\documentclass[journal, comsoc, twocolumn]{IEEEtran}
\pdfoutput=1 
\usepackage[utf8]{inputenc}
\usepackage{graphicx}
\usepackage{amsmath,amssymb}
\usepackage[urlcolor=black,hidelinks]{hyperref}
\usepackage{graphics,float,epsf}
\usepackage[justification=centering]{caption,subcaption}
\usepackage{epstopdf}
\usepackage{booktabs}
 \usepackage{algorithm,algorithmic}
\DeclareMathOperator*{\argmax}{argmax}
\usepackage{gensymb}
\usepackage{textcomp}
\usepackage[dvipsnames]{xcolor}
\usepackage[noadjust]{cite}
\usepackage{multirow}

\newcommand\Tstrut{\rule{0pt}{2.6ex}}         
\newcommand\Bstrut{\rule[-0.9ex]{0pt}{0pt}}   

\begin{document}
%

\title{Multi-User Provisioning in Millimeter-Wave\\Urban Cellular Networks}
%
%

\author{Aleksandar~Ichkov,~\IEEEmembership{}%
        Daniel~Sialkowski,~\IEEEmembership{}%
        Petri~M\"ah\"onen,~\IEEEmembership{}%
        and~Ljiljana~Simi\'c~\IEEEmembership{}
\thanks{All authors are with the Institute for Networked Systems, 
RWTH Aachen University, Aachen, Germany.}
\thanks{E-mail: \{aic, dsi, pma, lsi\}@inets.rwth-aachen.de}
\thanks{}
\thanks{"This work has been submitted to the IEEE for possible publication.  Copyright may be transferred without notice, after which this version may no longer be accessible."}
}

%

%



\maketitle

\begin{abstract}
In this paper we present the first comprehensive study of the multi-user capacity of millimeter-wave (mm-wave) urban cellular networks, using site-specific ray-tracing propagation data and realistic antenna array patterns. We compare the performance of TDMA and SDMA (time and spatial division multiple access, respectively) for diverse network scenarios and antenna configurations. We propose a greedy heuristic algorithm to solve the network-wide directional link allocation problem, thereby estimating the achievable capacity and coverage of multi-user mm-wave networks. Our results show that inter-cell interference is negligible, so that TDMA performance is strictly limited by air-time sharing. By contrast, the major limiting factor for SDMA is intra-cell interference, emphasizing the impact of real antenna array sidelobes. Nonetheless, SDMA significantly outperforms TDMA in terms of average UE throughput, by up to $2$~Gbps using $8\times8$ arrays. As an important design insight, our results show that larger base station antenna arrays limit intra-cell interference while compensating for small UE arrays, reducing costs and beamforming requirements in practical SDMA networks. Our analysis also shows that the limited number of antenna sub-arrays in a practical hybrid beamforming architecture may force SDMA to drop UEs with good coverage, highlighting a tradeoff between base station densification and antenna resources. 
\end{abstract}

\begin{IEEEkeywords}
Millimeter-wave, cellular networks, multi-user, TDMA, SDMA.
\end{IEEEkeywords}

%
\IEEEpeerreviewmaketitle

\section{Introduction}

	The spectrum-rich millimeter-wave (mm-wave) bands are a key candidate for solving the capacity crunch in future cellular networks~\cite{xiao_millimeter_2017}. The fundamental feasibility of outdoor mm-wave links has been demonstrated by a number of measurement studies~\cite{NYI_measurements, Ponttor_meas, weiler_measuring_2014}, and theoretical works have shown that high-capacity coverage can be achieved in urban areas given a sufficiently dense deployment of base stations (BS)~\cite{modeling_mmwSys, RStheory, Bai2014CC}. However, extending this great promise of high-capacity mm-wave \emph{links} to high-capacity \emph{networks} is predicated on effectively sharing the capacity among \emph{multiple} users in a cell. This is a more challenging task for mm-wave than in sub-$6$~GHz deployments, owing to the highly directional and blockage-prone nature of mm-wave links~\cite{NYI_measurements, Molisch2014_mmW}.

	In particular, intra-cell resource sharing among the associated UEs -- using a given multiple-user access control (MAC) scheme -- at mm-wave also requires active antenna beam management by the serving BS, unlike in conventional cellular deployments with static, wide-sector BS coverage. Namely, mm-wave requires either switching the direction of the BS antenna beam to serve UEs in different time slots using TDMA (time division multiple access), or forming multiple beams pointed at individual UEs to serve them simultaneously using SDMA (spatial division multiple access). The network-wide performance of these MAC schemes thus depends on the site-specific distribution of line-of-sight (LOS) or non-LOS (NLOS) mm-wave link opportunities~\cite{Secon2017}, as well as the spatial interference footprint of the antenna beams. Therefore, in evaluating the capacity of multi-user mm-wave cellular networks it is imperative to jointly consider the effects of MAC resource sharing and network-wide, spatially-dependent mm-wave coverage and interference. Yet, despite the intense research interest in mm-wave for mobile networks over the past decade (see e.g. ~\cite{xiao_millimeter_2017} and references therein), the network-wide performance of multi-user mm-wave cellular deployments has thus far been largely unaddressed in the literature.

	In this paper we present an extensive study of the achievable capacity of multi-user mm-wave cellular networks, for a wide range of network deployment scenarios and antenna configurations. We compare the performance of SDMA and TDMA in terms of the network coverage and UE throughput and explicitly characterize the corresponding distributions of both inter- and intra-cell interference over the network. In order to do so, we formulate the network-wide link allocation problem -- assigning to each UE the serving BS and the orientation of its directional antenna beam towards the UE -- and solve it using a greedy heuristic algorithm that aims to maximize the network throughput while serving as many UEs as possible. We consider realistic spatial distributions of mm-wave LOS/NLOS link opportunities over the network using 3D ray-tracing propagation data for two distinct urban sites in Frankfurt and Seoul, and capture the interference impact of antenna sidelobes by considering realistic antenna array patterns. Our results show that inter-cell interference is negligible, so that TDMA performance is strictly limited by air-time sharing. By contrast, intra-cell interference is the major limiting factor of SDMA performance, especially given the significant impact of real antenna array sidelobes. An important design insight from our analysis is that intra-cell interference can be limited using larger BS antenna arrays while compensating for small UE arrays, reducing both terminal costs and the overhead of mm-wave mobility management in practical SDMA networks.

	To the best of our knowledge, our work is the first to comprehensively study the multi-user capacity of TDMA and SDMA mm-wave urban cellular networks and characterize the corresponding distributions of inter- and intra-cell interference assuming realistic models of site-specific mm-wave propagation and directional antenna array patterns. The seminal works on mm-wave cellular network rate and coverage bounds in~\cite{RStheory, modeling_mmwSys} neglect any multi-user aspects, i.e. implicitly consider one UE per cell. Moreover, as is typical of theoretical analyses using stochastic geometry tools, these works assume statistical models of the mm-wave channel and urban blockage effects, and idealized sectored directional antenna models. The impact of realistic antenna array patterns on mm-wave network capacity is studied in~\cite{Zorzi}, but likewise considering statistical propagation models and only one UE per cell. On the other hand, a large number of works have focused on MAC design for mm-wave but typically \emph{within} a single cell, including signal processing aspects of beamforming e.g.~\cite{jointSpatialMultiplexing2014, Sun2014} and initial access and beam management schemes ~\cite{mmWMAC, Zorzi_3GPPbeam}. The authors in~\cite{LiCaireRRM, XueNonOrthBeams2017} are among the few to  study the problem of resource allocation in SDMA mm-wave systems, but only for a single cell. Moreover,~\cite{LiCaireRRM} focuses on concurrent transmission scheduling using a statistical channel model that neglects the spatial dependence of the intra-cell interference, whereas~\cite{XueNonOrthBeams2017} does model the interference in the angular domain but considers LOS links only (with a log-distance path loss model) and assumes idealized sectored antenna beams which, as our results show, severely limits the realism of intra-cell interference modeling for SDMA. Lastly, the authors in~\cite{Kularni2016-MIMOComparison, Bai2014CC} study the feasibility of multi-user MIMO with hybrid beamforming (HBF) versus single-user analog beamforming in mm-wave networks -- in MAC terms corresponding to SDMA and TDMA, respectively. However, these works also derive network capacity bounds by employing simplified channel and antenna models to aid analytical tractability, thereby failing to properly capture the spatially-dependent nature of MAC performance in mm-wave networks. 

	The rest of this paper is organized as follows. Sec.~\ref{sec:System_model} details our network, propagation and directional antenna models and MAC schemes studied. In Sec.~\ref{sec:Link_allocation} we present our link and interference models, and we formulate the multi-user network-wide link allocation problem and present our link allocation heuristic. Sec.~\ref{sec:Results} presents our results of multi-user mm-wave network performance. Sec.~\ref{sec:Conclusions} concludes the paper.

\section{System Model}
\label{sec:System_model}

\subsection{Network Model} \label{sec:network_model}

\begin{figure*}[t!]
	\centering
	\begin{subfigure}[]{0.5\textwidth}   	
          \centering
  	 \includegraphics[width=0.65\columnwidth]{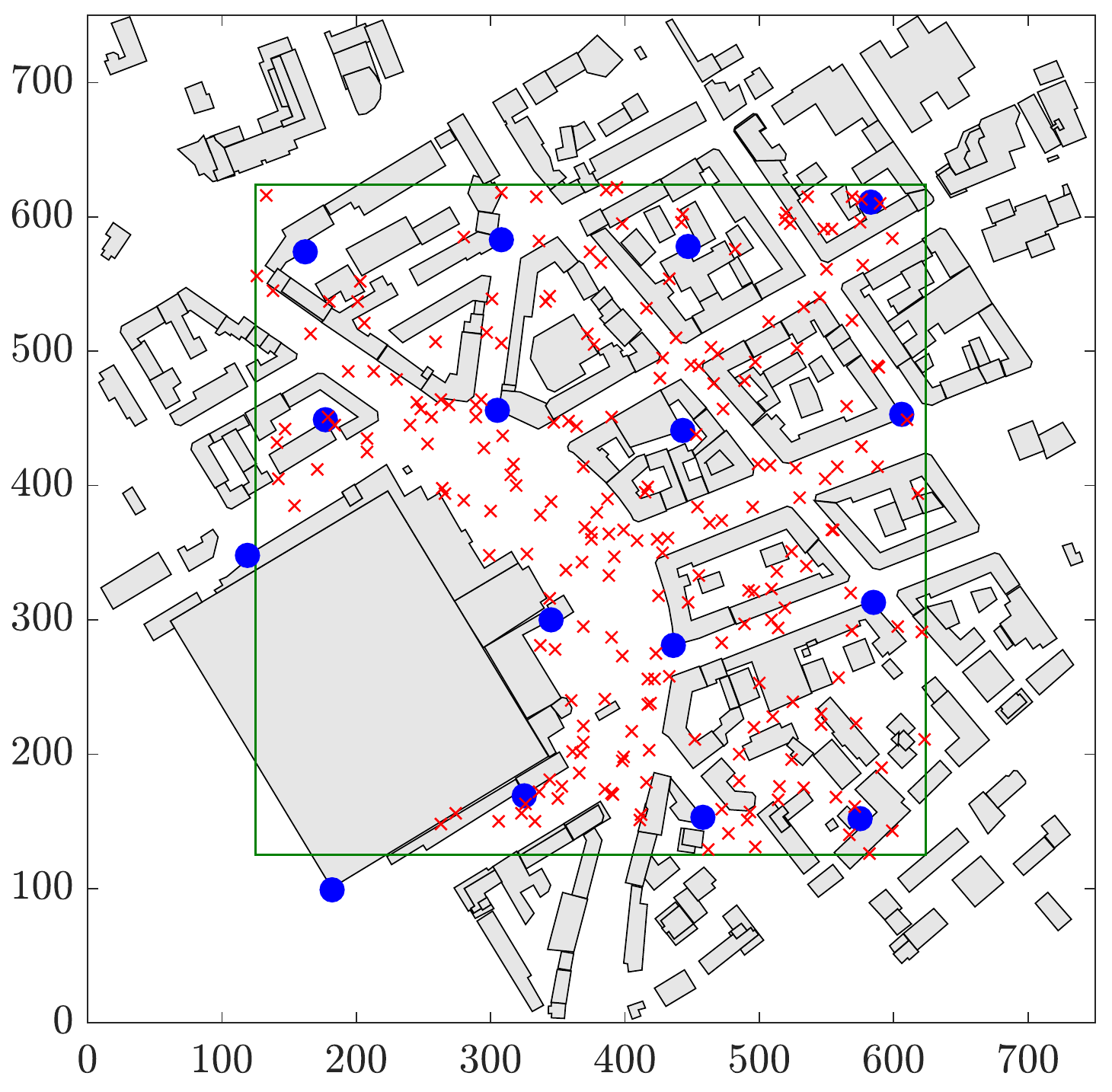}
  	 \caption{\textit{Frankfurt}}
	\end{subfigure}%
	~
	\begin{subfigure}[]{0.5\textwidth}   	
          \centering
   	\includegraphics[width=0.65\columnwidth]{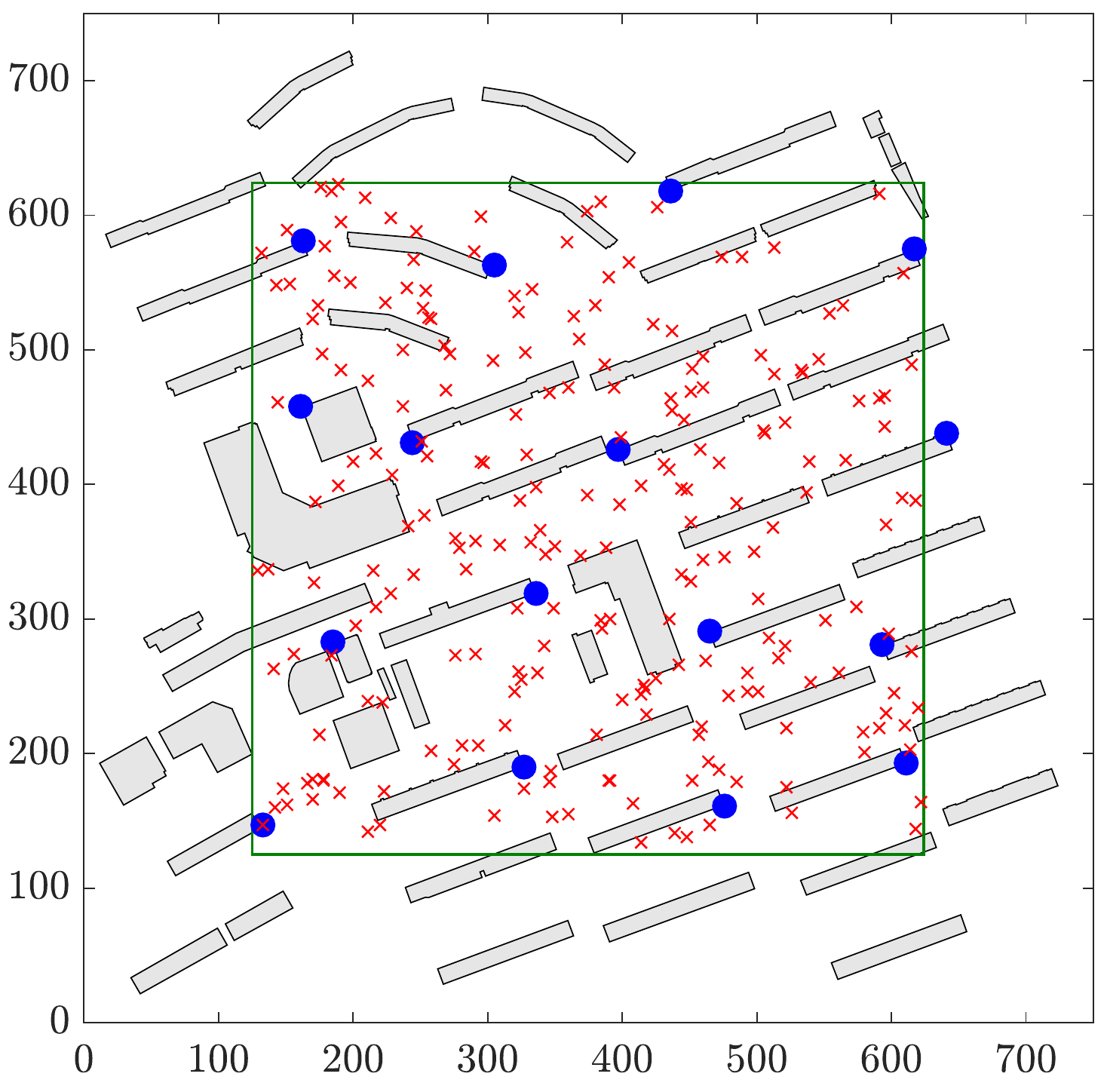}
   	\caption{\textit{Seoul}}
	\end{subfigure}
	\caption{Illustration of the urban network sites, showing the building layout and the BS (blue) and example UE (red) deployment over the study area (green box); $\lambda_{BS}=64$~$\text{BSs/km}^2$ and  $\lambda_{UE}=1000$~$\text{UEs/km}^2$.}
   	 \label{fig:0}
\end{figure*}

	We consider two distinct urban sites of $750$~$\text{m}$~$\times$~$750~\text{m}$ in Frankfurt and Seoul, in Fig.~\ref{fig:0}. The Frankfurt site is a densely built-up area near the main train station, typical of a busy central city, whereas the Seoul site is a more open-space residential area with wider streets and fewer buildings and street intersections. We consider a mm-wave cellular network consisting of BSs and users (UEs) that are deployed within a central $500$~$\text{m}$~$\times$~$500~\text{m}$ network area. The BS deployment follows a regular distribution on a uniform grid, where the BS locations are mapped to the closest building corner near the actual grid position. The BS locations are thus not optimized, but do represent a reasonable deployment layout (as building corners generally offer broad LOS coverage). We consider a range of BS densities, $\lambda_{BS}=\{32, 64, 100, 196\}$~$\text{BSs/km}^2$. The sparser deployments are representative of early mm-wave networks deployed as a capacity extension to legacy cellular networks, whereas the denser deployments correspond to future mm-wave stand-alone networks. We assume UEs uniformly randomly deployed over the network area following a Poisson point process (PPP) and consider a range of UE densities, $\lambda_{UE}=\{500, 1000, 1500\}$~$\text{UEs/km}^2$. 

		\begin{table}
			\centering
			\normalsize
			\begin{tabular}{l l}
				\toprule
				\textbf{Parameter} & \textbf{Value} \\ 
				\midrule 
				Urban sites & $\{$Frankfurt, Seoul$\}$\\ 
				Urban site size & $750$~$\text{m}$~$\times$~$750~\text{m}$\\
				Network area size & $500$~$\text{m}$~$\times$~$500~\text{m}$\\
				{BS} distribution & grid \\ 
				{BS} density, $\lambda_{BS}$ & $\{32, 64, 100, 196\}$~$\text{BSs/km}^2$\\ 
				{BS} height, $h_{BS}$ & $6$~$\text{m}$\\
				{UE} distribution, & $\{$uniform$\}$\\ 
				{UE} density, $\lambda_{UE}$ & $\{500, 1000, 1500\}$~$\text{UEs/km}^2$\\ 
				{UE} height, $h_{UE}$ & $1.5$~$\text{m}$\\
				\hline		
				Carrier frequency, $f_{c}$ & $60$~$\text{GHz}$ \Tstrut \\
				Max BS power, $P_{BS,max}$ & $30$~$\text{dBm}$\\
				Max {EIRP}, $EIRP_{max}$ & $40$~$\text{dBm}$\\ 
				Channel bandwidth, $B$ & $1$~$\text{GHz}$\\
				Receiver Noise Figure, $NF$ & $6$~$\text{dB}$\\			
				\bottomrule
		          	\end{tabular}
			\caption{Studied cellular network parameters.} 
			\label{table:systemParameters}
		\end{table}
		
	Frequency bands in the range of $24-86$~$\text{GHz}$ are under consideration for 5G-and-beyond cellular networks operating in the mm-wave bands. In this paper, we assume the network operates at the carrier frequency of  $f_{c}=60$~$\text{GHz}$ without loss of generality, and for comparability with a number of mm-wave outdoor urban measurements, e.g.~\cite{NYI_measurements, Molisch2014_mmW, weiler_measuring_2014, Ponttor_meas, imdea_60GHz}. We consider saturated downlink traffic and assume the BS transmits at the maximum power of $P_{BS,max}=30$~$\text{dBm}$, so as to respect the maximum EIRP of $EIRP_{max}=40$~$\text{dBm}$ (typical of spectrum regulatory limits, e.g. as set by FCC) given its specific antenna parameters (\emph{cf}.~Sections~\ref{sec:antennas}~and~\ref{sec:Interference_model}), with a bandwidth of $B=1$~GHz and receiver noise figure of $NF=6$~dB. We assume that BS and UE antennas are at a height of $h_{BS}=6$~$\text{m}$ and $h_{UE}=1.5$~$\text{m}$, respectively. Table~\ref{table:systemParameters} summarizes the studied network parameters. We assume the network is static, so that mobility-related MAC challenges such as initial user access \cite{Zorzi_IA_tutorial} or beam misalignment \cite{misalignment} are not considered. In practice, the network-wide aggregate throughput depends on real traffic and user mobility patterns, and user scheduling. Thus, our network model provides a first-order estimate of the achievable multi-user capacity.

\subsection{Propagation Model: mm-Wave Ray-Tracing}\label{sec:System_model_RAYTRACER}

	We use an open-source mm-wave ray-tracing tool~\cite{iNets_raytracer} to obtain site-specific propagation data, based on real 3D building data for the two urban study areas. We perform a dedicated ray-tracing simulation for each BS, considering a $1$~$\text{m}$~$\times$~$1~\text{m}$ UE receiver location grid over the study area, using a ray-launching angle granularity of $0.05^{\circ}$. Our ray-tracing tool considers free-space propagation and strong reflections (of up to two-bounces) as the dominant propagation mechanisms; diffraction is neglected, as it does not play a significant role at high frequencies~\cite{Molisch2014_mmW}. The final output is an inventory of all feasible propagation paths for all BS/UE pairs, giving for each propagation path $k$: (i) the nature of the path, LOS or NLOS; (ii) the angle of departure (AoD) and angle of arrival (AoA) at the BS and UE, $\{\phi_{BS, k},\theta_{BS, k}\}$ and $\{\phi_{UE, k},\theta_{UE, k}\}$, in the azimuth and elevation plane, respectively; and (iii) the path loss $L_k$, calculated as the free-space path loss along the propagation path plus any reflection losses. In order to consider all feasible strong reflected NLOS paths given the geometry of the urban layout, we consider the best-case scenario of all buildings having glass walls, assuming a loss of  $3$~$\text{dB}$ per reflection~\cite{Langen1994}. To capture all reflected paths relevant for the $500$~$\text{m}$~$\times$~$500~\text{m}$ network study area, we consider all buildings in the $750$~$\text{m}$~$\times$~$750~\text{m}$ site area in Fig.~\ref{fig:0}. 

\begin{figure*}[t]
	\centering
	\begin{subfigure}[t]{0.5\textwidth}
          \centering
  	 \includegraphics[width=0.7\columnwidth]{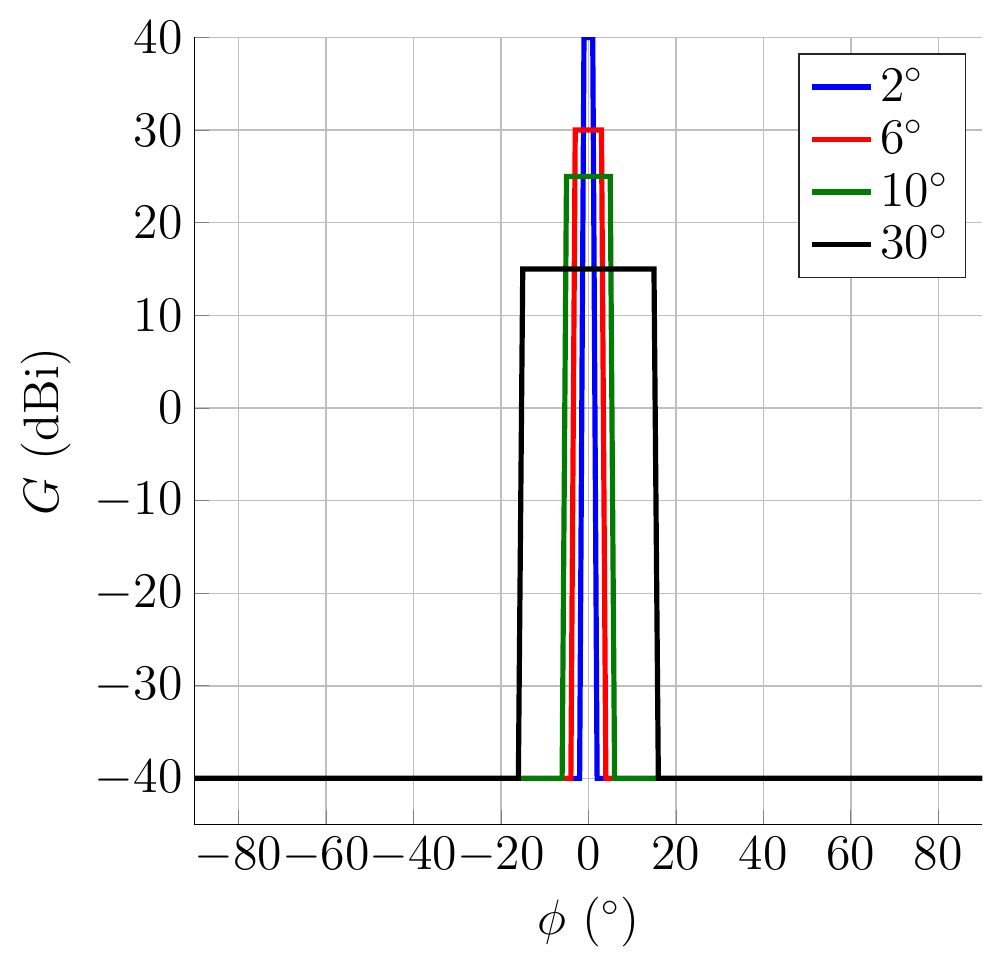}
  	 \caption{\textit{Ideal antennas}}
	\end{subfigure}%
	~	
	\begin{subfigure}[t]{0.5\textwidth}
          \centering
   	\includegraphics[width=0.7\columnwidth]{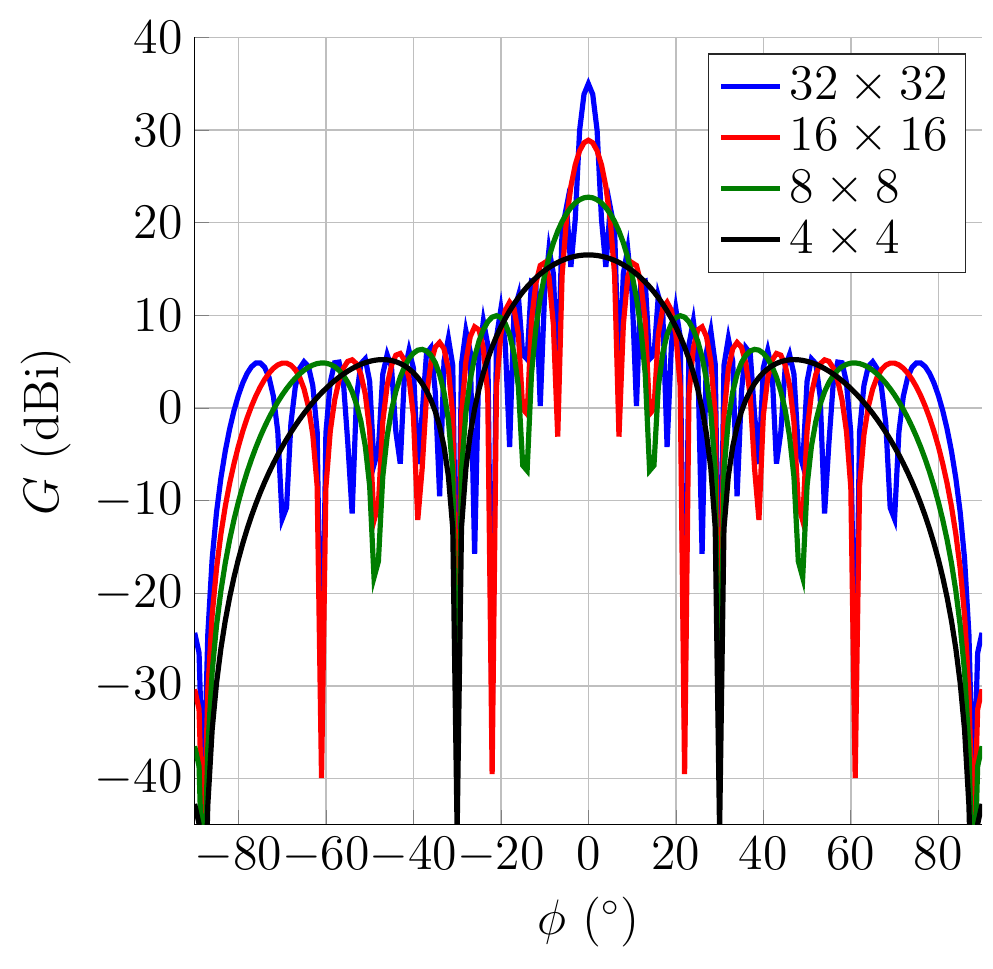}
  	 \caption{\textit{Realistic antenna arrays}}
	\end{subfigure}
   	 \caption{The considered antenna patterns in the azimuth plane.}
   	 \label{fig:Antennas}
\end{figure*}

	In general, this output of the omnidirectional ray-tracing simulations thus enables us to compute the link budget (and interference) for all feasible combinations of BS/UE link allocations over the network, for any transmit power allocation and for arbitrary (and arbitrarily oriented) directional antenna patterns assumed at the BS and UE, as detailed in Sec.~\ref{sec:Link_allocation}. If we restrict the orientations of the antennas to be aligned along a feasible propagation path $k$ between the BS and UE, to obtain the received signal strength (RSS) at the UE we simply post-process the ray-tracer output by aligning the main lobe pointing directions with the AoD and AoA of $k$ at the BS and UE, respectively. This corresponds to perfect beam alignment of the UE and its serving BS along the given LOS/NLOS path.

\subsection{Multi-User Access Schemes}
\label{sec:MAC}
	
	We consider two major MAC schemes: TDMA and SDMA. In TDMA, the BS uses time sharing to serve its associated UEs, allocating them individual timeslots. In SDMA, the BS instead uses spatial multiplexing to simultaneously serve multiple UEs using independent beams, with the total BS transmit power equally shared among all allocated links, as given by (\ref{equ:TX_power}). As a baseline, we also consider the single-user interference-free case, referred to as SU, which assumes independent user links by neglecting all interference effects, such that the UE obtains the maximum throughput achievable according to the signal-to-noise ratio (SNR). SU thus represents the upper bound of the network performance, i.e. as if the serving BS has zero cell load and all BS resources are available for serving a given UE. Considering the SU case alongside TDMA and SDMA thus allows us to differentiate between inherent propagation based mm-wave coverage for a single user and the additional effects of inter and intra-cell interference in a multi-user network (\emph{cf.}~Sec.~\ref{sec:Interference_model}).

\subsection{Directional Antenna and Beamforming Model} \label{sec:antennas}

	We consider two antenna types, ideal and realistic array antennas, of equivalent directivity, as specified in Table~\ref{table:Antennas} and shown in the azimuth-plane in Fig.~\ref{fig:Antennas}. This allows us to study in Sec.~\ref{sec:Results} the interference impact of modeling realistically the sidelobes of directional antennas in multi-user mm-wave networks. The ideal directional antenna is modeled by a sectored 2D pattern, with a constant maximum main lobe gain for the whole half-power beamwidth (HPBW) and a gain of \mbox{-40}~dBi otherwise; the tool AMan in the radio simulator WinProp was used to interpolate the corresponding 3D antenna pattern using horizontal projection interpolation. We generate realistic antenna array patterns assuming 2D uniform quadratic arrays with isotropic antennas as the basic array element, using the \mbox{MATLAB Phased Antenna System Toolbox}.

	We assume an idealized beamforming model where the main lobes of the UE and its serving BS are always perfectly beamformed towards the direction of the serving propagation path $k$ -- or interchangeably antenna orientation $k$ -- represented by the quadruple $\{\phi_{BS, k},~\theta_{BS, k},~\phi_{UE, k},~\theta_{UE, k}\}$, and that the nominal antenna pattern does not change with the steering direction. The antenna gains of the BS/UE along antenna orientation $k$ are thus given by	  
	 	\begin{equation} \label{equ:BSaligmentgain}
	G_{BS, k}=  G(\phi_{BS, k}-\phi_{BS, k},\theta_{BS, k}-\theta_{BS, k}) = G(0^\circ, 0^\circ),
	\end{equation}
	\begin{equation} \label{equ:UEaligmentgain}
	\resizebox{0.91\hsize}{!}{%
	 $G_{UE, k}=  G(\phi_{UE, k}-\phi_{UE, k},\theta_{UE, k}-\theta_{UE, k}) = G(0^\circ, 0^\circ).$%
	 }
	\end{equation}

	In general, the BS/UE antenna gains in the direction of an arbitrary propagation path $j$ between the BS and the UE, when their respective main lobes are steered towards the propagation path $k$ orientation, are given by
	
	\begin{equation} \label{equ:BSmisaligmentgain}
	G_{BS, k, j}=  G(\phi_{BS, k}-\phi_{BS, j},\theta_{BS, k}-\theta_{BS, j}),
	\end{equation}
	\begin{equation} \label{equ:UEmisaligmentgain}
	G_{UE, k, j}=  G(\phi_{UE, k}-\phi_{UE, j},\theta_{UE, k}-\theta_{UE, j}).
	\end{equation}
	
\begin{figure}[t!]
          \centering
  	 \includegraphics[scale=0.48]{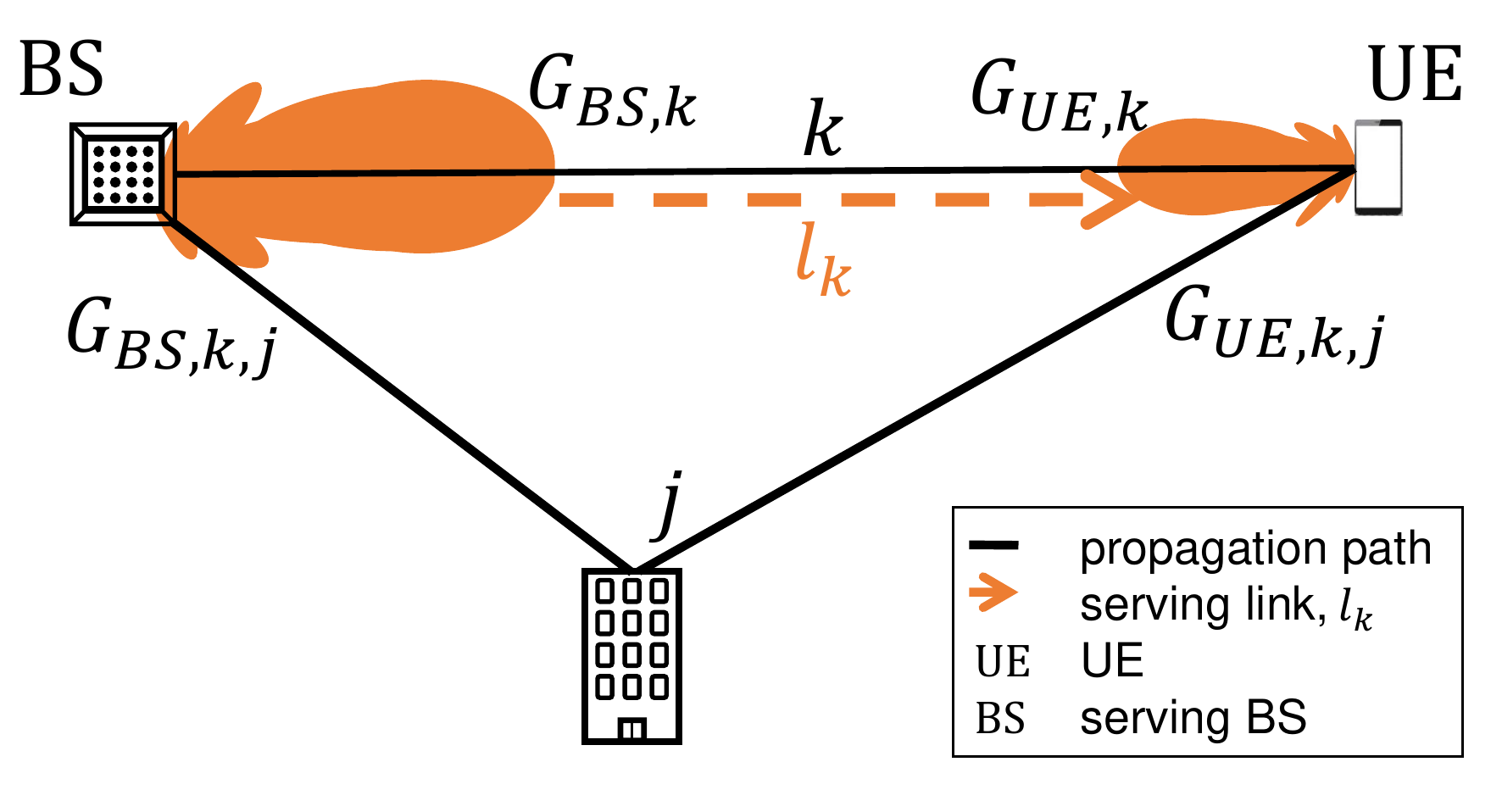}
   	 \caption{Illustration of the beamforming model, showing the BS/UE antennas gains in the direction of LOS propagation path $k$ and NLOS propagation path $j$, when the BS/UE main lobes are aligned with propagation path $k$,~\emph{cf}.~(\ref{equ:BSaligmentgain})--(\ref{equ:UEmisaligmentgain}).}
   	 \label{fig:servingLink}
\end{figure}

Fig.~\ref{fig:servingLink} illustrates the beamforming model, representing the BS/UE antennas gains when their main lobes are aligned towards the propagation path $k$, as given by (\ref{equ:BSaligmentgain})--(\ref{equ:UEmisaligmentgain}). 	
	
		\begin{table}[t!]
			\centering
			\normalsize
			\begin{tabular}{l l l l}
					\toprule
					\textbf{Antenna} & \textbf{\# of ant.} & \textbf{HPBW} & \textbf{Max gain} \textbf{(dBi)} \\ 
					\textbf{type} & \textbf{elements} & \textbf{(\textdegree)} & $G(0^\circ, 0^\circ)$\\ 
					\midrule
					Isotropic (ISO) & - & - &  0 \Bstrut\\
					\hline
					Ideal 30\textdegree & - & 30  & 15 \Tstrut \\ 
					Ideal 10\textdegree & - & 10  & 25 \\ 
					Ideal 6\textdegree & - & 6  & 30 \\ 
					Ideal 2\textdegree & - & 2  & 40 \Bstrut\\
					\hline
					Array 4$\times$4  & 16 & 26 &  16.5 \Tstrut\\ 
					Array 8$\times$8 & 64 & 12.4  & 22.8\\ 
					Array 16$\times$16 & 256 & 6 & 28.9\\ 
					Array 32$\times$32 & 1024 & 2.8 & 35\\ 
					\bottomrule
				\end{tabular} 
			\caption{Antenna configuration characteristics.}
			\label{table:Antennas}
		\vspace{-5pt}
		\end{table}

	In the multi-user network context, this corresponds to analog beamforming with TDMA, where the antenna orientations are fixed for a given BS-UE pair and the corresponding timeslot duration. For SDMA, we assume HBF, where each UE is served using an individual beam, as formed by a separate BS antenna sub-array. The sub-array is steered to the antenna orientation along the serving propagation path $k$ for a given BS-UE pair, ignoring any other simultaneously served UEs. Namely, no special beamshaping is applied for interference mitigation, such that UEs may experience intra-cell interference. Similarly, we assume no additional beamshaping for inter-cell coordination, such that UEs may experience inter-cell interference, for both TDMA and SDMA. We quantify both types of interference throughout, as defined in Sec.~\ref{sec:Interference_model}.

\section{Multi-User mm-Wave Link Allocation}
\label{sec:Link_allocation}

In this section, we present our link and interference models, and we formulate the link allocation problem and present our proposed heuristic solution.
	
\subsection{Link Model}
\label{sec:link_model}

	A mm-wave link $l_k$ is defined by the serving BS, the associated UE, their respective antenna orientations, $\{\phi_{BS, k},~\theta_{BS, k}\}$ and $\{\phi_{UE, k},~\theta_{UE, k}\}$, the signal-to-interference-plus-noise ratio (SINR) $SINR_{l_k}$ and the corresponding achievable throughput $R_{l_k}$. We note that the serving link $l_k$, along the propagation path $k$, is allocated to the UE as per the link allocation strategy, and that $SINR_{l_k}$ and $R_{l_k}$ depend on the network-wide link allocation (\emph{cf}.~Sec.~\ref{sec:link_problem}). 
	
	The total gain $T_{l_k}$ on link $l_k$ can be expressed as the total gain over all feasible propagation paths, \textbf{N}$_{{p, BS, UE}}$, between the serving BS and the UE, assuming that the BS and the UE main lobes are steered in the direction of the propagation path $k$, and is given by  

	\begin{equation}  \label{equ:T_lk}
	T_{l_k} = \sum\limits_{\substack{j=1}}^{N_{p, BS, UE}} {\frac{G_{BS, k, j} \cdot G_{UE, k, j}} {L_j}}, 
	\end{equation}

\noindent where $L_j$ denotes the path loss of the propagation path~$j$, and $G_{BS, k, j}$, $G_{UE, k, j}$ are given by (\ref{equ:BSmisaligmentgain}),~(\ref{equ:UEmisaligmentgain}), respectively.
		
	The corresponding SINR of link $l_k$ can be expressed as 
	\begin{equation} \label{equ:SINR}
	\text{SINR}_{l_k} =  \frac{T_{l_k} \cdot P_{BS, l_k}} {I_{intra,l_k} + I_{inter, l_k} + N},
	\end{equation}

\noindent where $I_{intra, l_k}$ and $I_{inter, l_k}$ denote the intra-cell and inter-cell interference, as given by (\ref{equ:intra}) and (\ref{equ:inter}), respectively, and detailed in \ref{sec:Interference_model}, and $N$ is the noise power, consisting of the thermal noise over $B$ plus $NF$. $P_{BS,l_k}$ denotes the BS transmit power allocated to link~$l_k$,   
	
\begin{eqnarray}
\small
\label{equ:TX_power}
P_{BS,l_k} = \begin{cases} 
\dfrac{P_{BS,max}}{N_{l,BS}} & \mbox{if } M < EIRP_{max}, \\
\dfrac{EIRP_{max} - G_{BS,max}}{N_{l,BS}} & \mbox{otherwise}, 
\end{cases}
\end{eqnarray}

\noindent where $M=P_{BS,max}+G_{BS,max}$ and $N_{l,BS}$ is the number of allocated beams (links) at the BS for SDMA; for TDMA, $N_{l,BS}=1$. 

	We estimate the UE throughput of link $l_k$ using a truncated Shannon bound model \cite{etsi_lte}, as
\begin{eqnarray}
\small
\label{equ:LTEthroughput}
R_{l_k} = \begin{cases} 
0 &  SINR_{l_k} < SINR_{min}, \\
a_r R_{max} &  SINR_{l_k} > SINR_{max}, \\ 
a_r \alpha B \log_2 (1+SINR) & \mbox{otherwise},
\end{cases}
\end{eqnarray}
where $\alpha=0.6$ represents implementation losses, ${SINR}_{min} = -10$~dB is the minimum SINR for coverage, ${SINR}_{max} = 22.05$~dB is SINR for the maximum throughput of $R_{max} = 4.4$~Gbps, and $a_r$ is the air-time ratio, $a_r=\frac{1}{N_{l, BS}}$ for TDMA and $a_r=1$ for SDMA.

\subsection{Interference Modeling} 
\label{sec:Interference_model}	

	We distinguish between inter-cell and intra-cell interference. In TDMA networks, UEs associated with the BS are served in individual timeslots, so only inter-cell interference is considered. In SDMA networks where all UEs are served simultaneously\footnote{Unless stated otherwise, we assume an idealized HBF architecture where each BS is equipped with a sufficient (unlimited) number of antenna sub-arrays, so that enough beams can be generated to simultaneously serve all associated UEs.}, both intra-cell and inter-cell interference must be taken into account. 
	
\begin{figure}[t]
          \centering
  	 \includegraphics[scale=0.345]{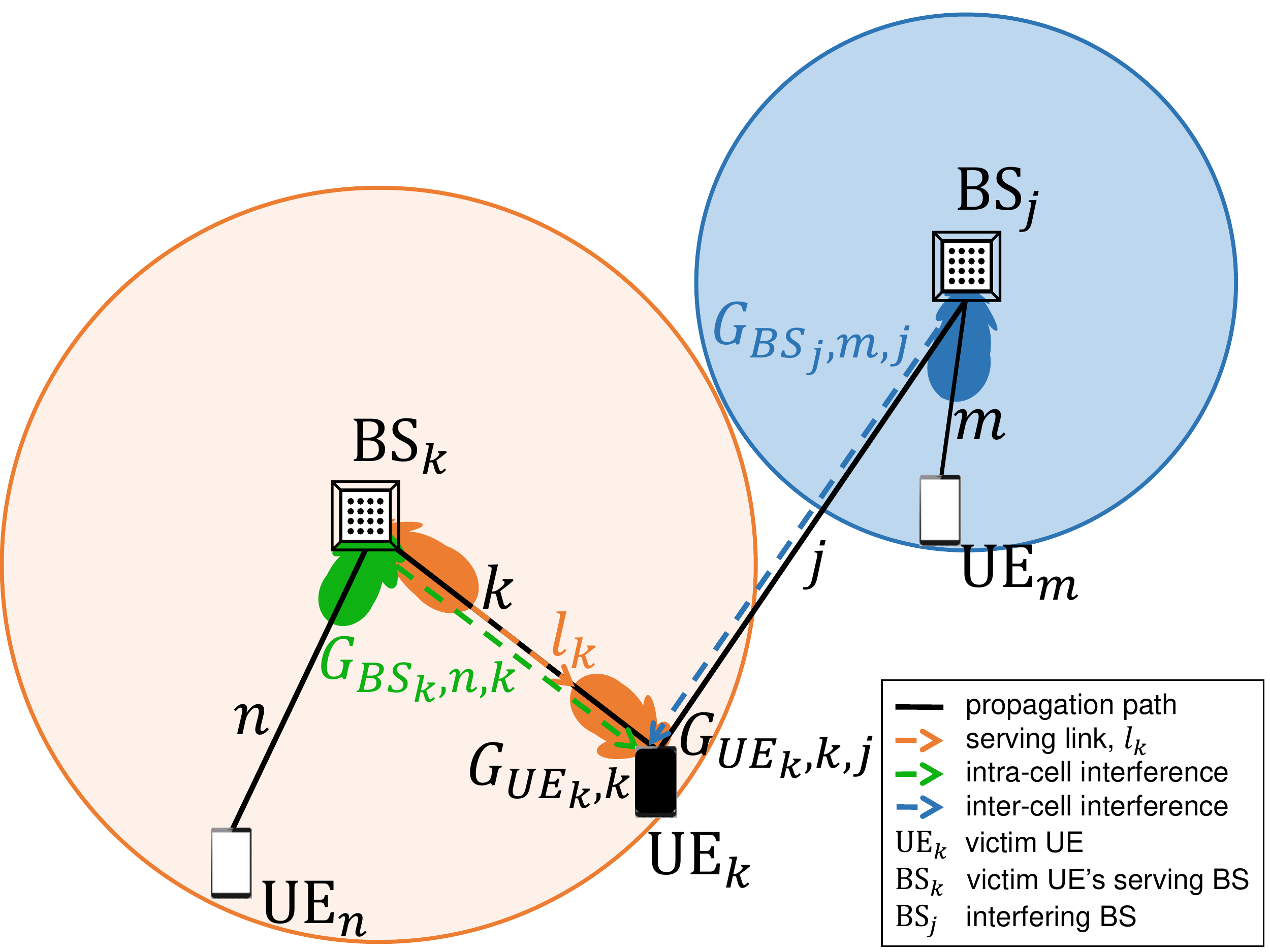}
   	 \caption{Illustration of the interference model for a victim link~$l_k$, where UE$_k$ is served by BS$_k$ and receives $I_{intra}$ from link~$l_n$ in the direction of the serving propagation path~$k$ and $I_{inter}$ from link~$l_m$ and interfering BS$_j$ in the direction of the propagation path~$j$.} 
   	 \label{fig:interferingLink}
\end{figure}

	The \textit{intra-cell interference} for a victim link~$l_k$ can be calculated as the aggregate interference from all other links allocated on the serving BS (\textbf{N}$_{{l, BS}}\setminus l_k$) and over all feasible propagation paths between the serving BS and the victim UE (\textbf{N}$_{{p, BS, UE}}$): 	
			\begin{equation} \label{equ:intra}
			I_{intra,l_k} = \sum\limits_{\substack{l_i=1 \\ l_i \neq l_k }}^{N_{l, BS}} \sum\limits_{\substack{j=1}}^{N_{p, BS, UE}} \frac{G_{BS, i, j} \cdot G_{UE, k, j} \cdot P_{BS, l_i}}{L_j},
			\end{equation}	

\noindent where $P_{BS, l_i}$ denotes the BS transmit power allocated to the interfering link $l_i$, as given by (\ref{equ:TX_power}), $G_{BS, i, j}$ denotes the BS antenna gain in the direction of the propagation path $j$ when the BS main lobe is steered in the direction of the interfering propagation path~$i$, as given by (\ref{equ:BSmisaligmentgain}) and $G_{UE, k, j}$ denotes the UE antenna gain in the direction of the propagation path~$j$ when the UE main lobe is steered in the direction of the serving propagation path $k$, as given by (\ref{equ:UEmisaligmentgain}).
	
	The \textit{inter-cell interference} for a victim link~$l_k$ can be calculated as the aggregate interference from all interfering BSs (\textbf{N}$_{BS}\setminus BS_k$) and all links allocated on each interfering BS (\textbf{N}$_{{l, BS}}$) and over all feasible propagation paths between each interfering BS and the victim UE (\textbf{N}$_{{p, BS, UE}}$):
			\begin{equation} \label{equ:inter}\resizebox{0.88\hsize}{!}{%
	 $I_{inter,l_k} = \sum\limits_{\substack{BS=1 \\ BS \neq BS_{k}}}^{N_{BS}} \sum\limits_{\substack{l_i=1}}^{N_{l, BS}} \sum\limits_{\substack{j=1}}^{N_{p, BS, UE}} \frac{G_{BS, i, j} G_{UE, k, j} P_{BS, l_i} a_r}{L_j }$},%
			\end{equation}

\noindent where, the probability that an interfering link $l_i$ will be active at the same time as the serving victim link $l_k$ is given by the air-time ratio $a_r=\frac{1}{N_{l, BS}}$ for TDMA; for SDMA, $a_r=1$.

	The interference model for an example victim link~$l_k$ is illustrated at Fig.~\ref{fig:interferingLink}, showing the corresponding BS and UE antenna gains for the interfering links $l_n$~(intra-cell interference for SDMA) and $l_m$~(inter-cell interference, for SDMA and TDMA).

\subsection{Link Allocation Problem}
\label{sec:link_problem}

	The link allocation problem for our multi-user mm-wave network consists of deciding, for each UE in the network, which BS to associate with and which antenna orientations the UE and its BS should use. The network-wide link allocation should maximize the network throughput while ensuring as many UEs are served as possible. Since there are an infinite number of different antenna orientations, there are in theory an infinite number of mm-wave links between a BS and UE. However, given the spatially sparse nature of mm-wave channels, we limit the set of potential links to those corresponding to antenna orientations along feasible propagation paths. The solution of the link allocation problem will thus be an element from the finite set of combinations of all feasible links.
	
	Assuming that a UE can only be served by one BS using one link at a time, the link allocation problem can be formulated as

	\begin{equation} \label{equ:alloc}
	\textbf{L}_{{allocation}}=\argmax_{l_k \in N_{l}} \; \sum\limits_{\substack{BS=1}}^{N_{BS}} \sum\limits_{\substack{l_k=1}}^{N_{l, BS}} R_{l_k}, 
	\end{equation}
\noindent s.t.
	\begin{equation} \label{equ:coverageConstraint}
	SINR_{l_k} \geq SINR_{min},  
	\end{equation}
	\begin{equation} \label{equ:TX_Power}
	\sum\limits_{\substack{l_k=1}}^{N_{l, BS}} P_{BS, l_k} \leq P_{BS,max}, 
	\end{equation}
	\begin{equation} \label{equ:numUEsConstraint}
	N_{l, BS} \leq N_{limit}, 
	\end{equation}
	
\noindent where $\textbf{L}_{{allocation}}$ denotes the network-wide link allocation set and \textbf{N}$_{l}=\bigcup\limits_{UE=1}^{N_{UE}} \bigcup\limits_{BS=1}^{N_{BS}} \textbf{N}_{{p,BS,UE}}$ denotes the set of all feasible links for all UEs over all BSs. The constraint in (\ref{equ:coverageConstraint}) states the coverage threshold for each link allocation; (\ref{equ:TX_Power}) states the maximum BS transmit power limit for the allocated links; (\ref{equ:numUEsConstraint}) ensures that the number of allocated links per BS cannot exceed $N_{limit}$, the number of BS sub-arrays (in SDMA).

\subsection{Link Allocation Heuristic}
\label{sec:Link_allocation_heuristic}

	The combinatorial link allocation optimization problem in (\ref{equ:alloc}) can be solved by evaluating all potential allocations via exhaustive search, but this is computationally infeasible for all but very small networks. Therefore, we consider a heuristic approach for the link allocation based on the greedy algorithm\footnote{Although our heuristic may in general provide a non-optimal solution, for small networks we have confirmed  that it is near-optimal.} in Algorithm~\ref{alg:heuristic} which adds candidate links sequentially, checking whether the constraints (\ref{equ:coverageConstraint})-(\ref{equ:numUEsConstraint}) are met following each allocation, to create the final network-wide link allocation set.
	
		\newfloat{algorithm}{tb!}{lop}
		\begin{algorithm}
			\caption{Link allocation heuristic}
			\label{alg:heuristic}
			\small
			\renewcommand{\algorithmicrequire}{\textbf{Input:}}
			\renewcommand{\algorithmicensure}{\textbf{Output:}}
			\begin{algorithmic} [1]
				\REQUIRE {
					        $N_{limit}$, maximum number of UEs that the BS can serve. \\
					        \textbf{N}$_{l}$, set of all feasible links for all UEs over all BSs.} 
				\STATE \textbf{Initialization:} 	\\
									\textbf{L}$_{{allocation}}=\{\}$, set of allocated links. \\
 									\textbf{c}$_l$ = sort($\textbf{N}_{l}$), 
						  			set of all candidate links for all UEs, sorted in descending order of the throughput $R$ (\ref{equ:LTEthroughput}) by assuming SINR=SNR.\\
						  			$bs_c$, serving BS for the candidate link $l_c$. \\ 
									$N_{l, bs_c}=0$, number of allocated links on the BS $bs_c$.
				\WHILE  {is not empty \textbf{c}$_l$}	
					\STATE $l_c \leftarrow \textbf{c}_l$, candidate link with highest achievable throughput for an arbitrary UE $ue_c$. \\ 
					\IF {$(SINR_{l_c} \geq SINR_{min}$ AND $N_{l,bs_c} < N_{limit})$}
						\STATE \textbf{L}$_{\text{allocation}} \leftarrow l_c$,  add candidate link $l_c$ to allocated links. \\
							 Update all allocated links performance metrics: \\
							 Re-calculate $SINR_{l_k}, R_{l_k}$, $\forall l_k \in\textbf{L}_{{allocation}}$.\\
							\IF {$(SINR_{l_k} \geq SINR_{min}$, $\forall l_k \in \textbf{L}_{{allocation}}$)}
							\STATE $N_{l, bs_c} = N_{l, bs_c}+1$, update the number of allocated links on the BS $bs_c$.  \\
								  \textbf{c}$_l = \textbf{c}_l \setminus \textbf{c}_l(ue_c)$, remove all candidate links for UE $ue_c$. 
							\ELSE
						           \STATE \textbf{L}$_{{allocation}}= \textbf{L}_{{allocation}} \setminus l_c$, remove $l_c$ from the link allocation set. \\
							\textbf{c}$_l = \textbf{c}_l \setminus l_c$, remove $l_c$ from candidate links set. \\
							 Update all allocated links performance metrics: \\
							 Re-calculate $SINR_{l_k}, R_{l_k}$, $\forall l_k \in\textbf{L}_{{allocation}}$.\\
						           \ENDIF
					\ELSE
					\STATE \textbf{c}$_l = \textbf{c}_l \setminus l_c$, remove $l_c$ from the candidate links set.
					\ENDIF
				\ENDWHILE
				\ENSURE  {\textbf{L}$_{{allocation}}$, final link allocation set.}
			\end{algorithmic}
		\end{algorithm}
	
	The algorithm uses the set of all feasible links $\textbf{N}_{l}$ as an input. A set of all candidate links \textbf{c}$_l$ for all UEs is created, sorted in descending order of the throughput. Each candidate link $l_c$ is then checked for whether it satisfies the minimum SINR threshold constraint~(\ref{equ:coverageConstraint}) and whether there are available resources at the candidate serving BS for a new link allocation~(\ref{equ:numUEsConstraint}). If so, $l_c$ is provisionally added to the link allocation set \textbf{L}$_{{allocation}}$ and the performance metrics for all previously allocated links are updated accordingly (SINR and throughput). If the updated performance metrics for all allocated links still satisfy the constraint in (\ref{equ:coverageConstraint}), the UE candidate link $l_c$ is permanently allocated to \textbf{L}$_{{allocation}}$ and the number of allocated links on the serving BS, $N_{l, bs_c}$ is incremented. Before proceeding further, all candidate links for that UE are removed from \textbf{c}$_l$. Otherwise, if $l_c$ does not satisfy (\ref{equ:coverageConstraint}), if there are no available resources at $bs_c$, or if allocating $l_c$ degrades the SINR of any of the previously allocated links below $SINR_{min}$, $l_c$ is removed from \textbf{c}$_l$ and the performance metrics of allocated links in \textbf{L}$_{{allocation}}$ are reverted to their previous values. The algorithm then proceeds with the next candidate link from \textbf{c}$_l$ for a given UE. After the complete set \textbf{c}$_l$ is exhausted, we get the final link allocation set \textbf{L}$_{{allocation}}$ as an output. If none of the candidate links for a given UE satisfies the constraints, the UE will not be served and will be dropped from the network.

\section{Results}
\label{sec:Results}

In this section we present our results of multi-user mm-wave network performance, analysing the effect of: (i) MAC scheme, (ii) network density, (iii) antenna configuration and (iv) HBF with a limited number of sub-arrays. Throughout, the results represent Monte Carlo simulations with five network realizations of the random UE deployments. 

\subsection{Basic Performance of MAC Schemes}
\label{sec:baseline}

\begin{figure}[t]
	\centering
	\begin{subfigure}[]{0.5\textwidth}	
	\includegraphics[width=0.95\columnwidth]{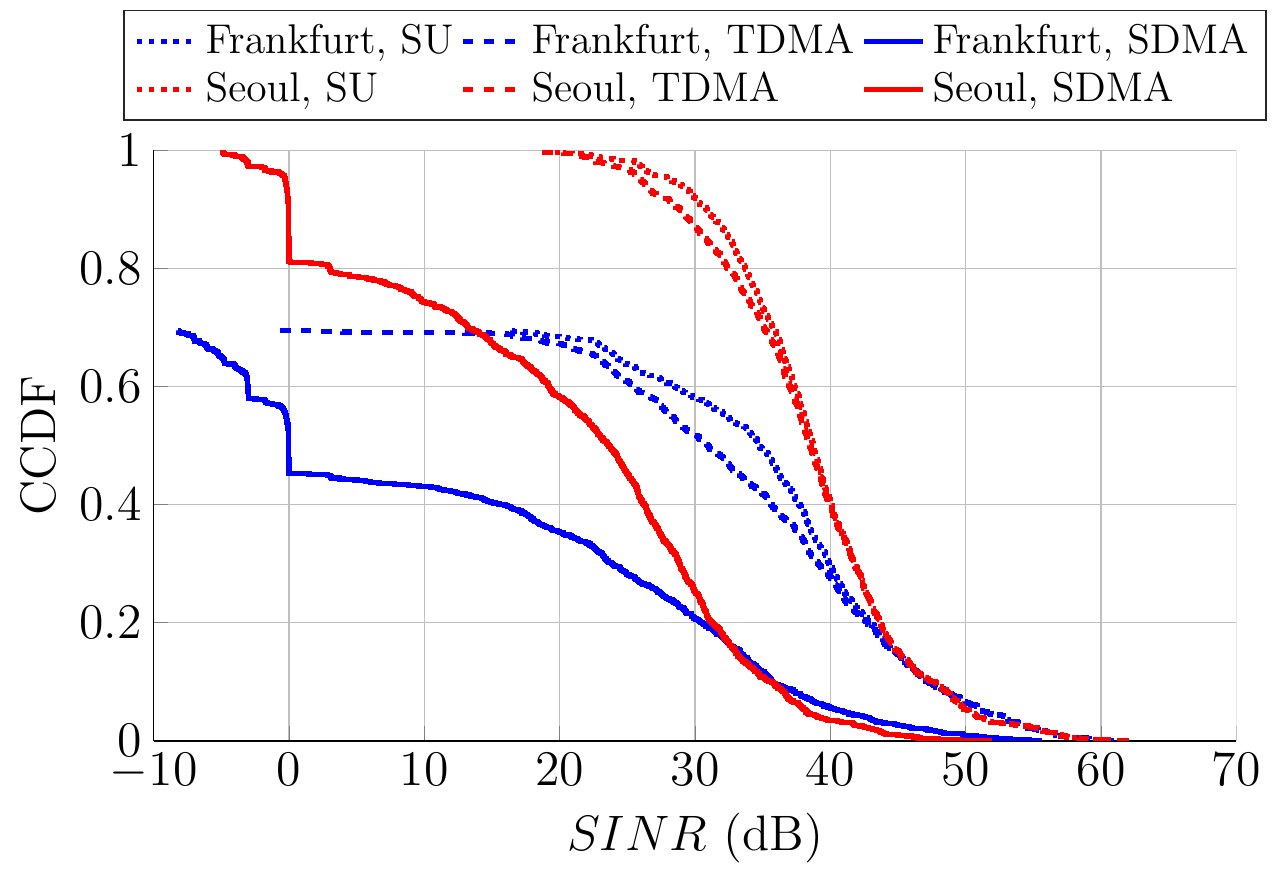}
  	 \caption{\textit{}}
  	 \label{fig:1a}
	\end{subfigure}%
	
	\begin{subfigure}[]{0.5\textwidth}	
	\includegraphics[width=0.95\columnwidth]{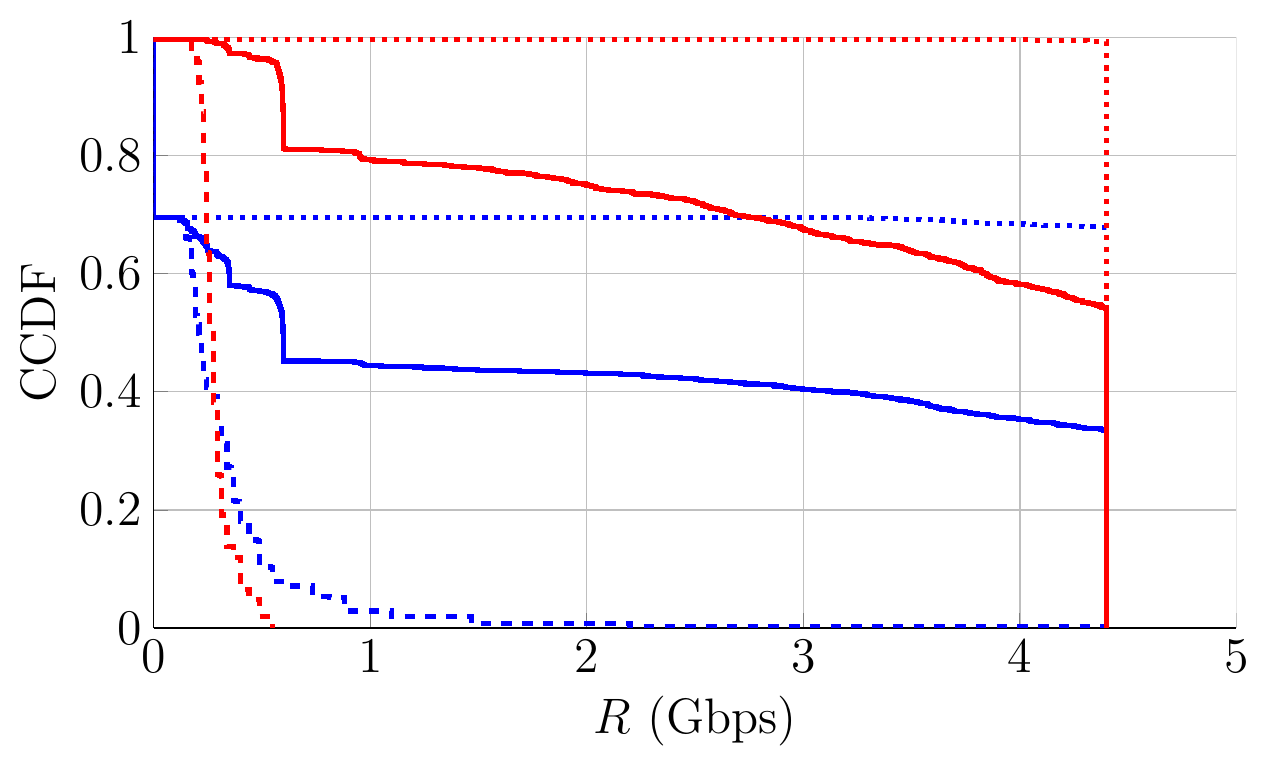}
   	\caption{\textit{}}
   	\label{fig:1b}
	\end{subfigure}

   	 \caption{SINR and throughput distributions showing the effect of MAC schemes ($\lambda_{BS}=64$~$\text{BSs/km}^2$, $\lambda_{UE}=1000$~$\text{UEs/km}^2$, $10^{\circ}$ ideal antennas).}
   	 \label{fig:1}
\end{figure}

\begin{figure}[h!]
	\includegraphics[width=0.95\columnwidth]{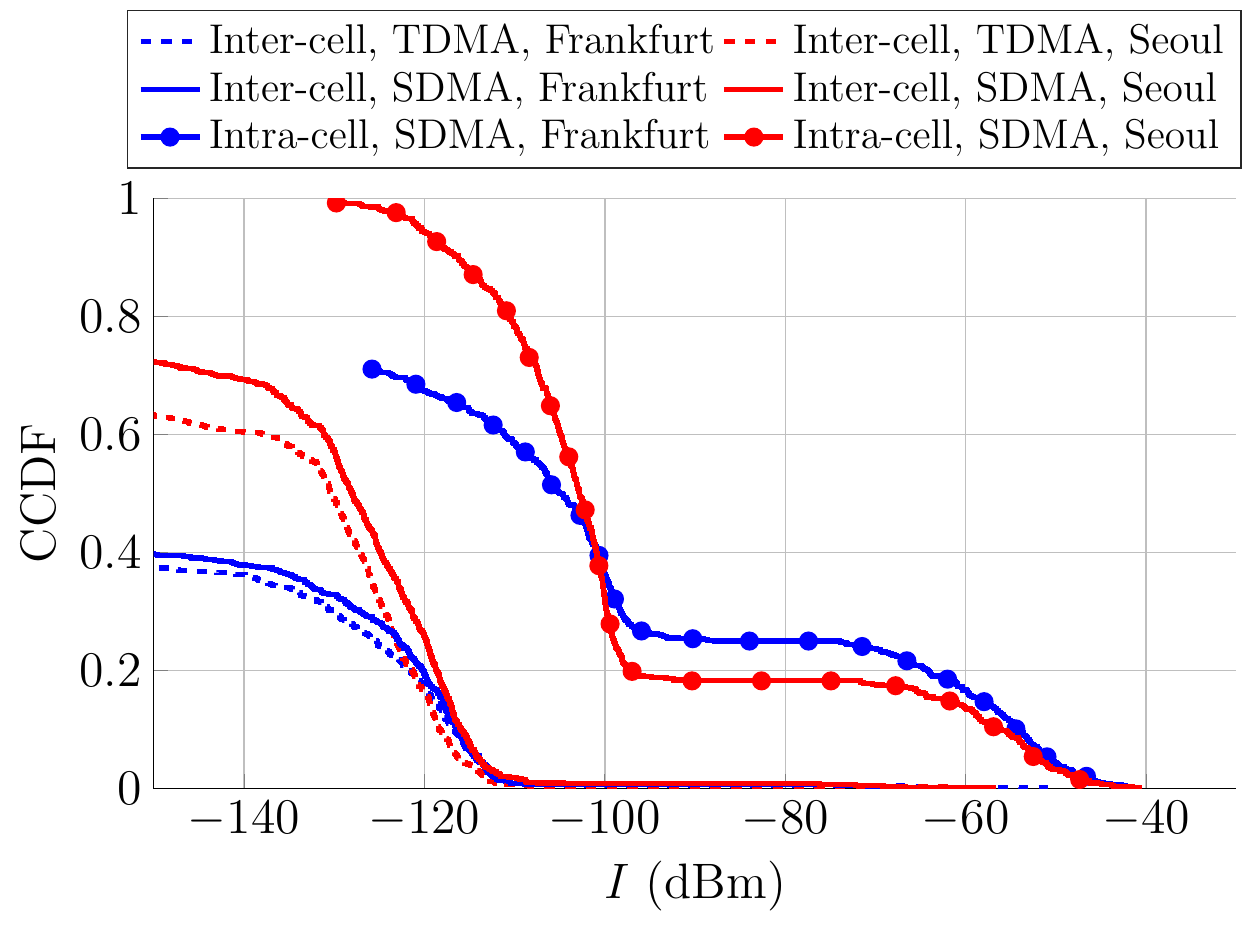}
	\caption{Interference component distributions showing the effect of MAC schemes ($\lambda_{BS}=64$~$\text{BSs/km}^2$, $\lambda_{UE}=1000$~$\text{UEs/km}^2$, $10^{\circ}$ ideal antennas).}
  	  \label{fig:3}
\end{figure}
	
\begin{figure*}[h!]
    \centering
    \begin{subfigure}[]{0.29\textwidth}
           \hspace{-0.15cm}
	\includegraphics[scale=0.46]{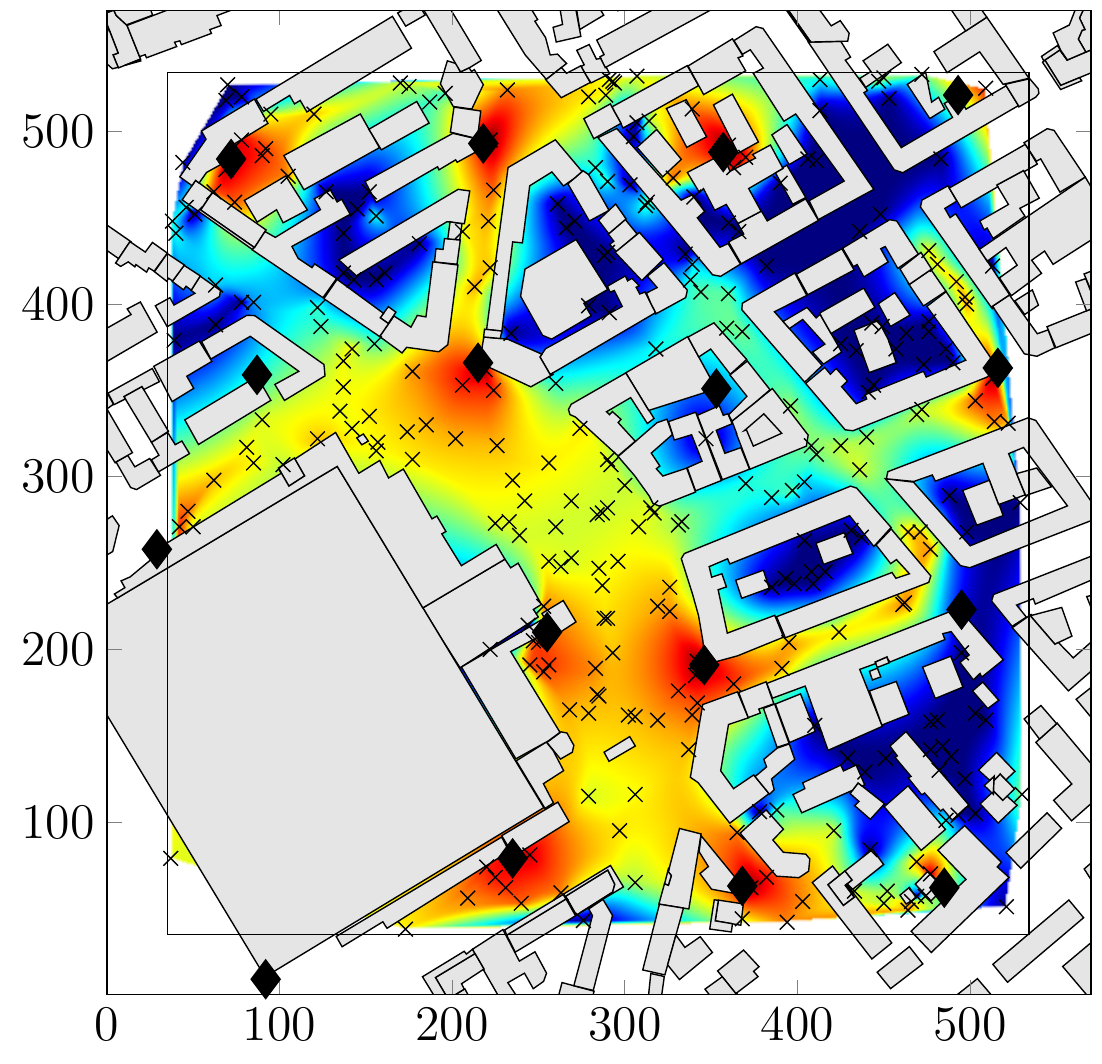}
	\caption{\textit{SU, SINR}}
  	  \label{fig:Frankfurt_heatmaps_a}
    \end{subfigure}%
     ~
    \begin{subfigure}[]{0.29\textwidth}
           \hspace{-0.15cm}
	\includegraphics[scale=0.46]{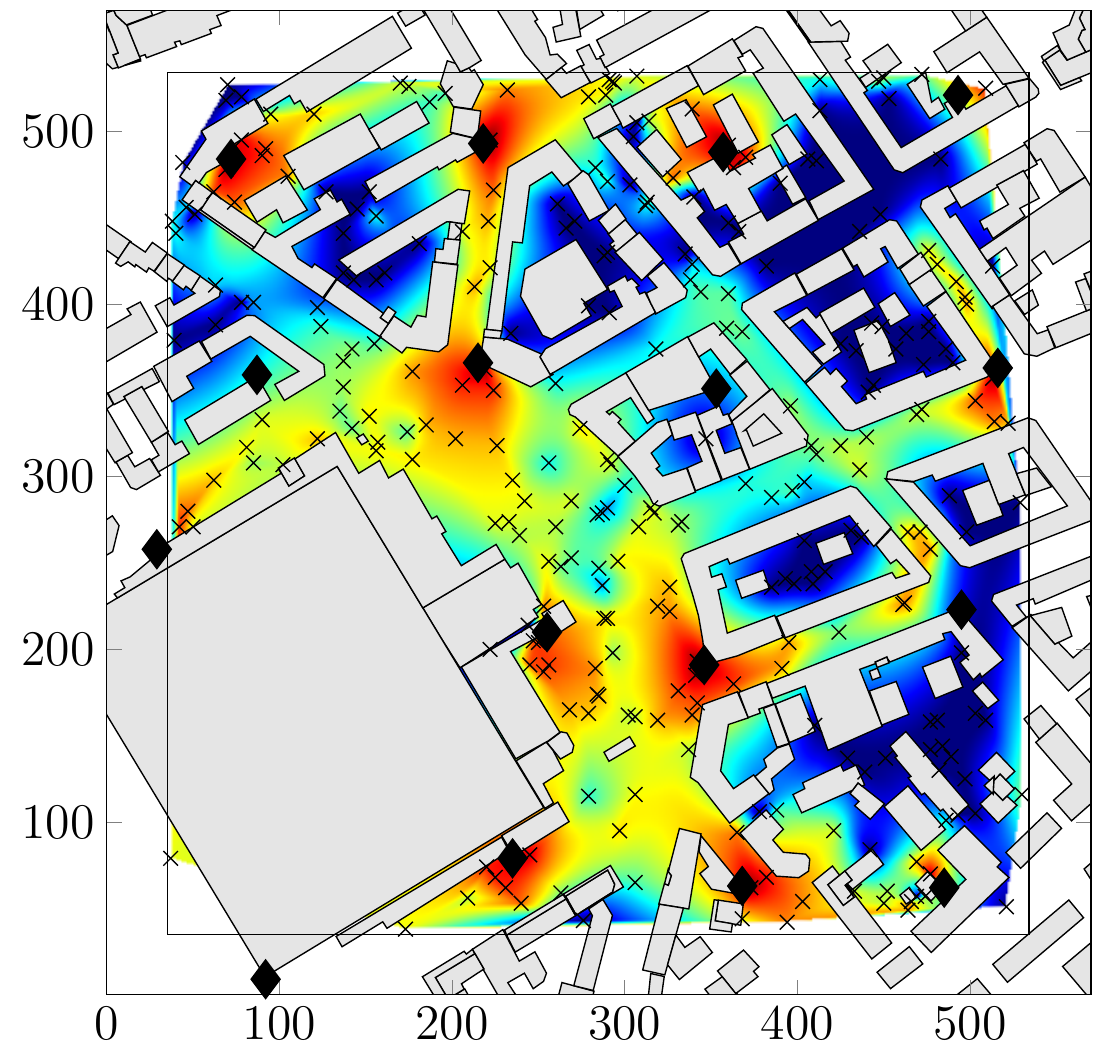}
	\caption{ \textit{TDMA, SINR}}
  	  \label{fig:Frankfurt_heatmaps_b}
    \end{subfigure}
    ~
    \begin{subfigure}[]{0.29\textwidth}
           \hspace{-0.1cm}
	\includegraphics[scale=0.46]{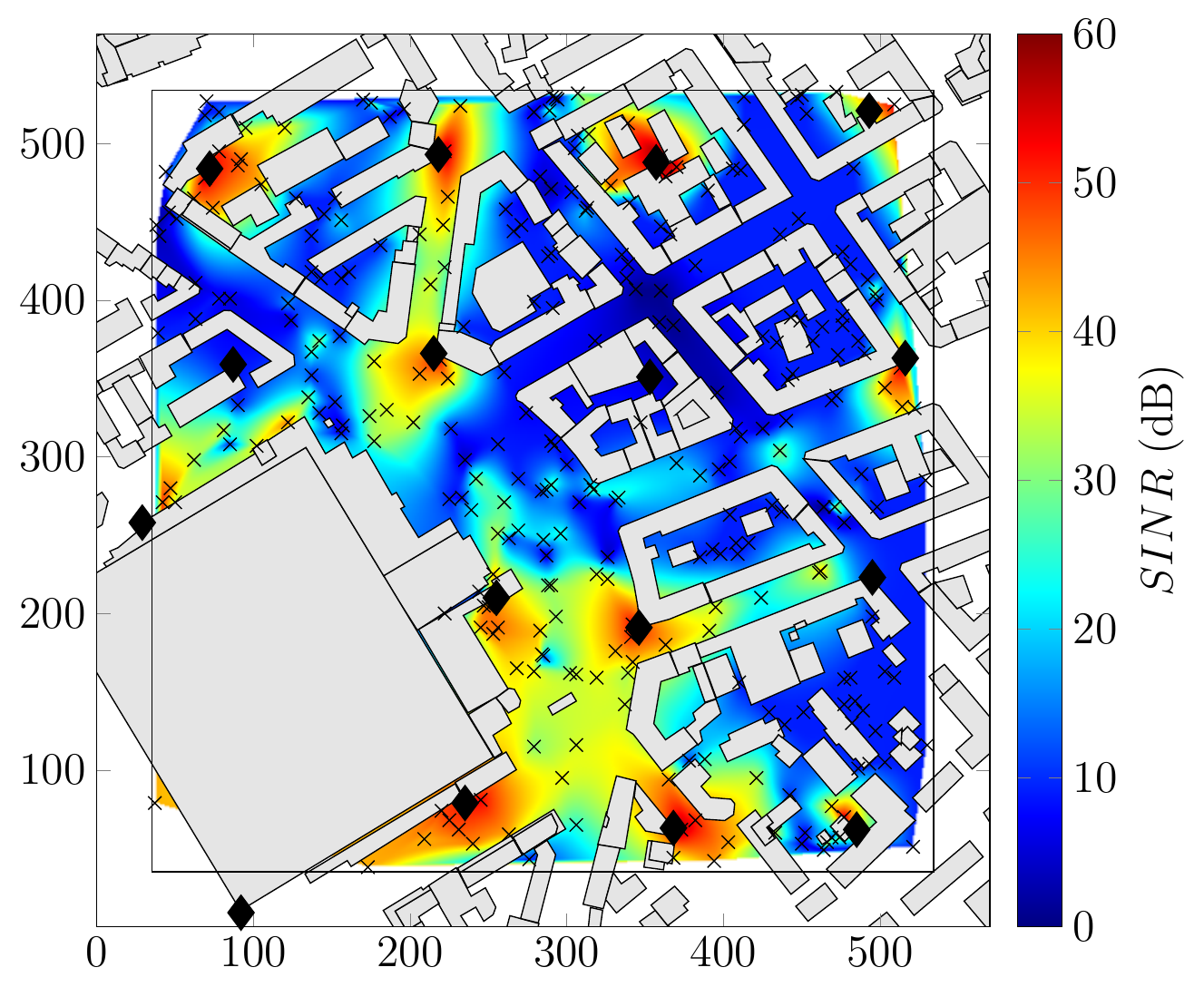}
	\caption{ \textit{SDMA, SINR}}
  	  \label{fig:Frankfurt_heatmaps_c}	
    \end{subfigure}
    \begin{subfigure}[]{0.29\textwidth}
           \hspace{-0.15cm}
	\includegraphics[scale=0.46]{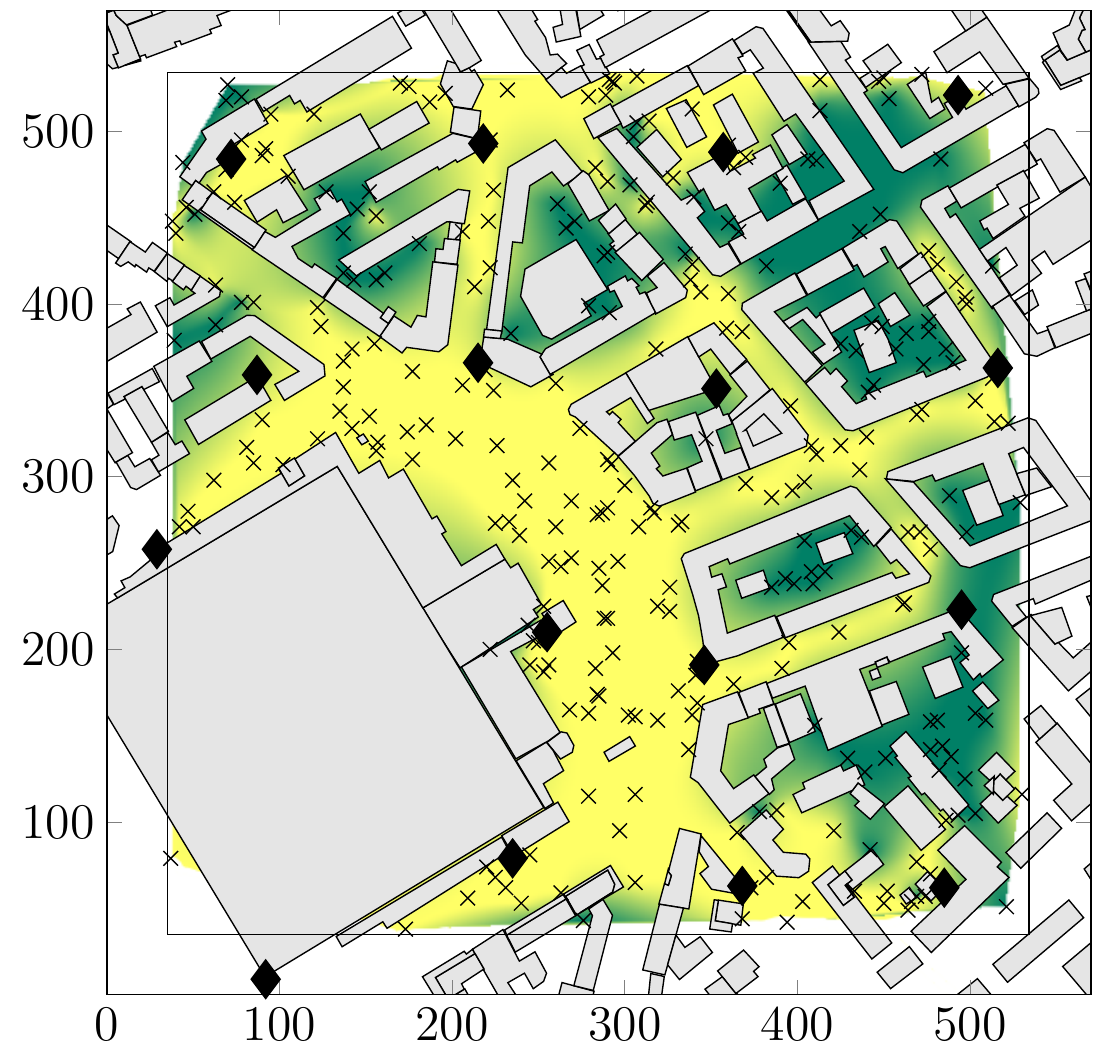}
	\caption{\textit{SU, UE throughput}}
	 \label{fig:Frankfurt_heatmaps_d}
    \end{subfigure}%
     ~
    \begin{subfigure}[]{0.29\textwidth}
           \hspace{-0.15cm}
	\includegraphics[scale=0.46]{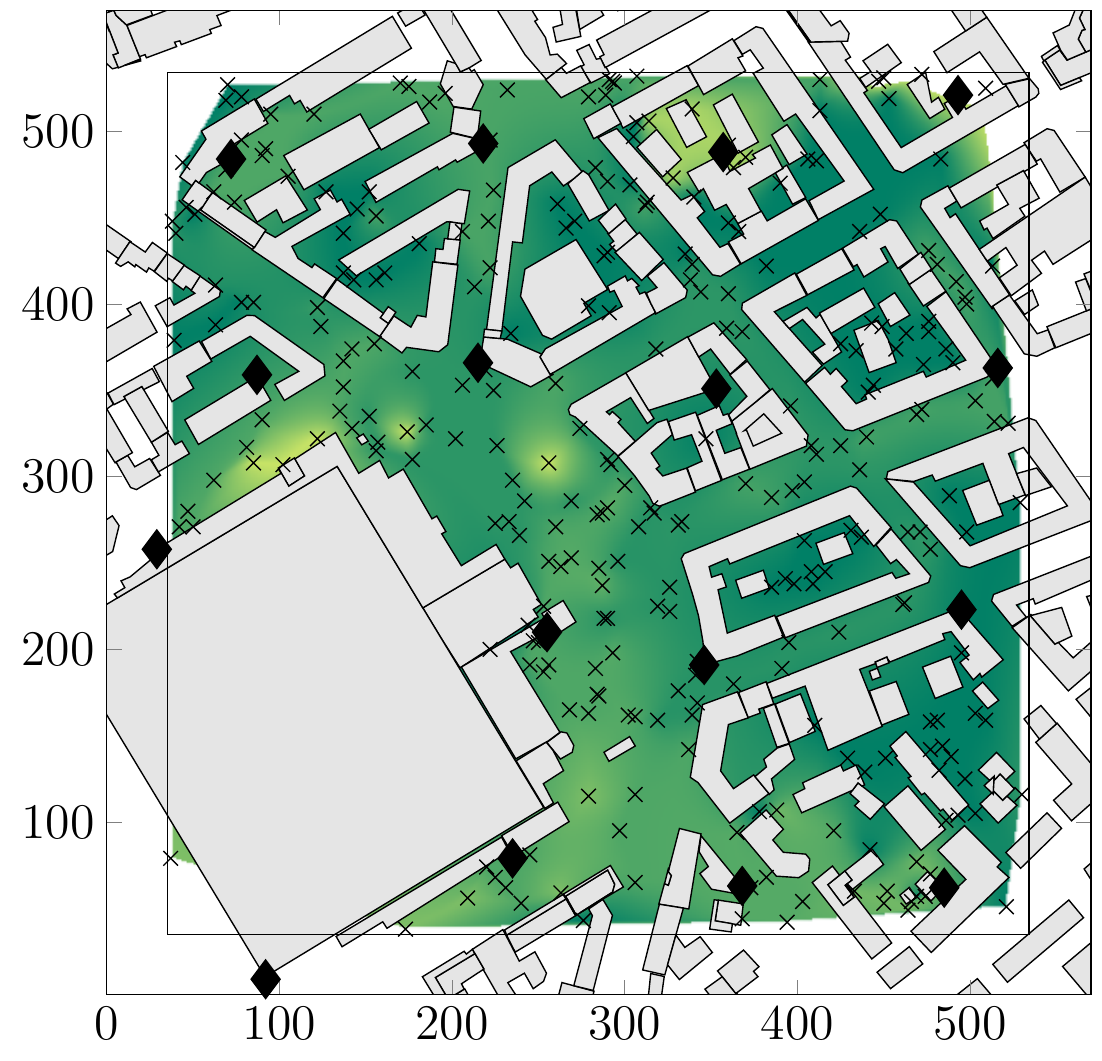}
	\caption{ \textit{TDMA, UE throughput}}
	\label{fig:Frankfurt_heatmaps_e}
    \end{subfigure}
    ~
    \begin{subfigure}[]{0.29\textwidth}
           \hspace{-0.1cm}
	\includegraphics[scale=0.46]{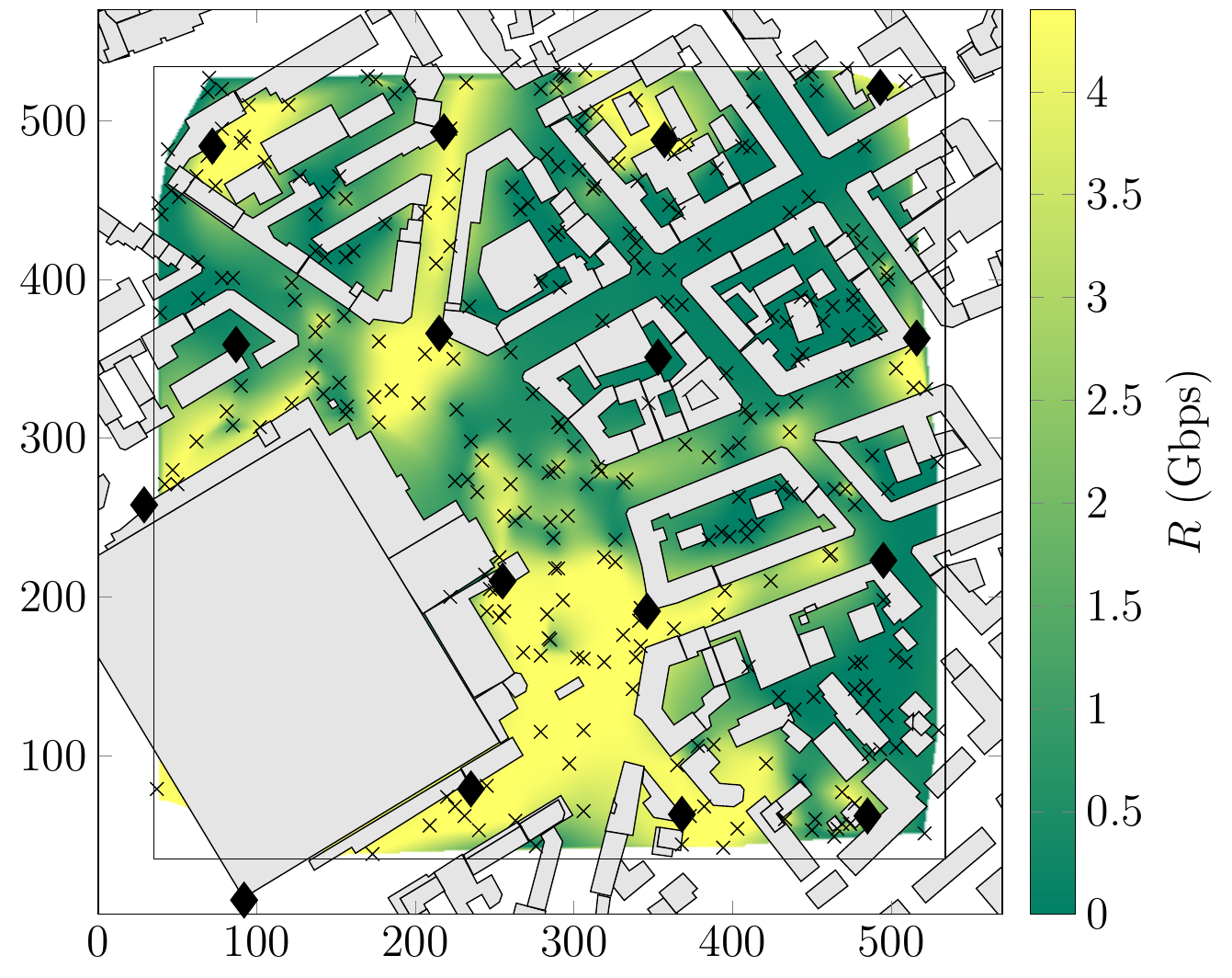}
	\caption{ \textit{SDMA, UE throughput}}
	  \label{fig:Frankfurt_heatmaps_f}
    \end{subfigure}
        \begin{subfigure}[]{0.29\textwidth}
           \hspace{-0.15cm}
	\includegraphics[scale=0.46]{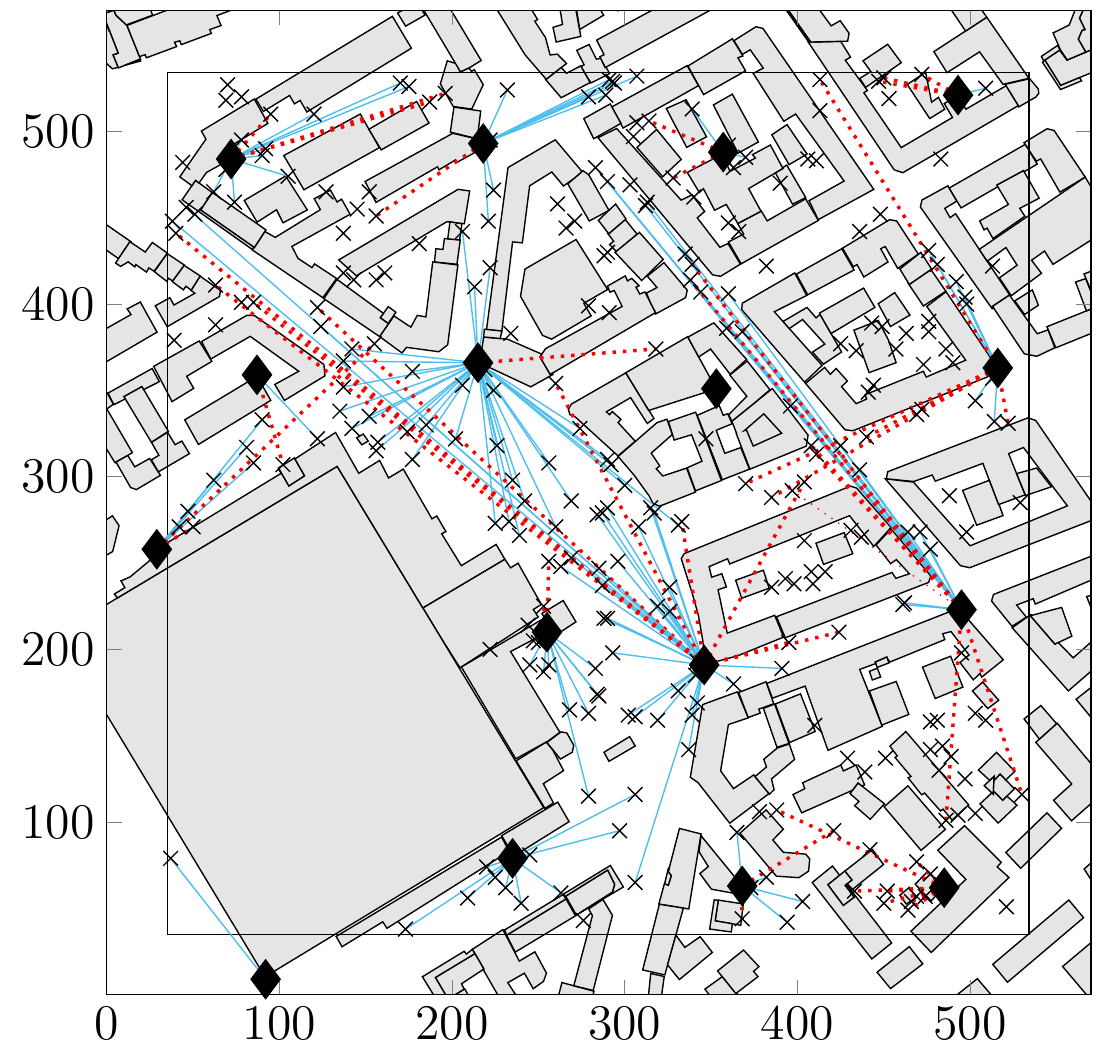}
	\caption{\textit{SU, allocated links, \\ LOS ($\textcolor{cyan}{\rule[0.5ex]{0.4cm}{0.48pt}}$) and NLOS ($\textcolor{red}{--}$)}}
	 \label{fig:Frankfurt_heatmaps_g}
    \end{subfigure}%
     ~
    \begin{subfigure}[]{0.29\textwidth}
           \hspace{-0.15cm}
	\includegraphics[scale=0.46]{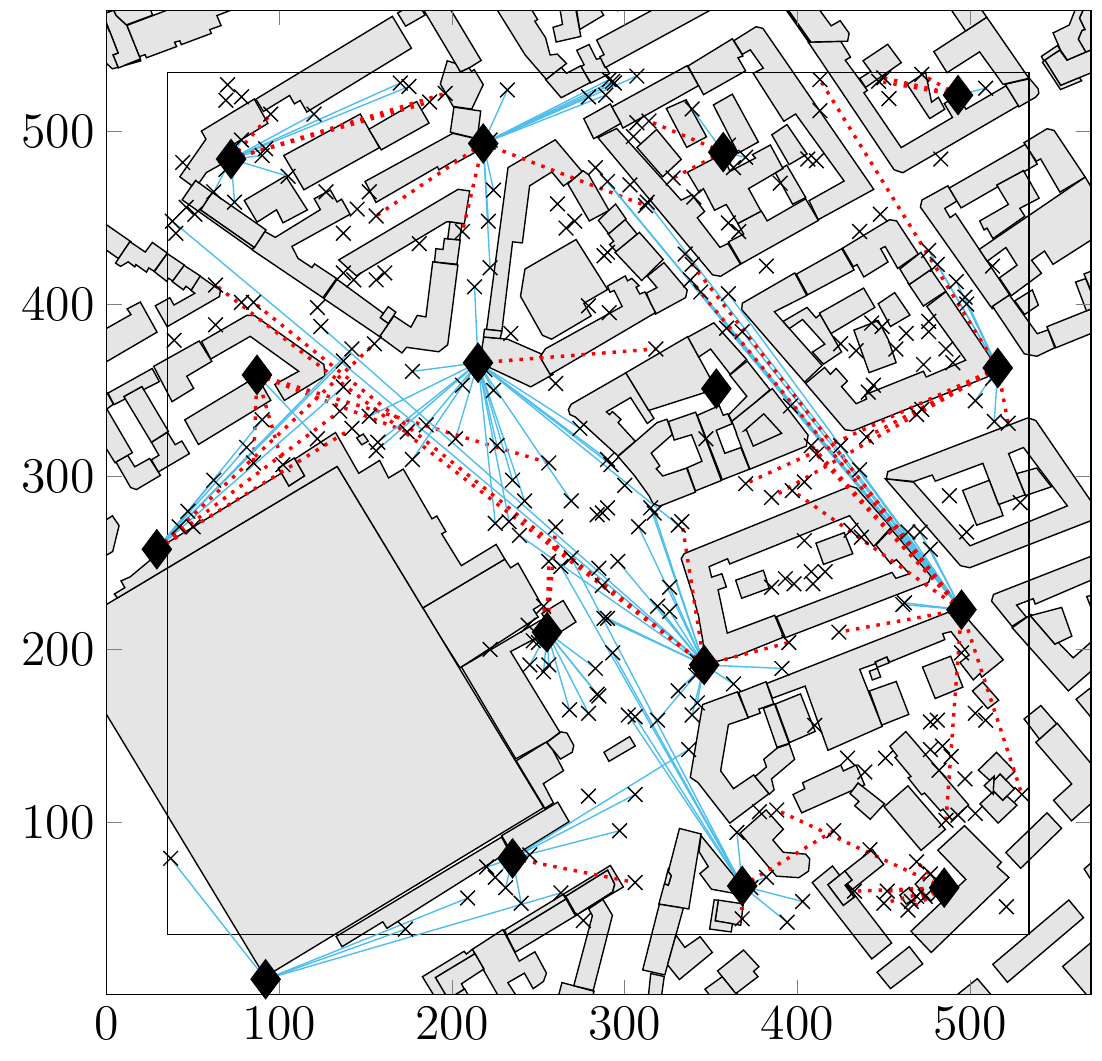}
	\caption{ \textit{TDMA, allocated links, \\ LOS ($\textcolor{cyan}{\rule[0.5ex]{0.4cm}{0.48pt}}$) and NLOS ($\textcolor{red}{--}$)}}
  	  \label{fig:Frankfurt_heatmaps_h}
    \end{subfigure}
    ~
    \begin{subfigure}[]{0.29\textwidth}
           \hspace{-0.15cm}
	\includegraphics[scale=0.46]{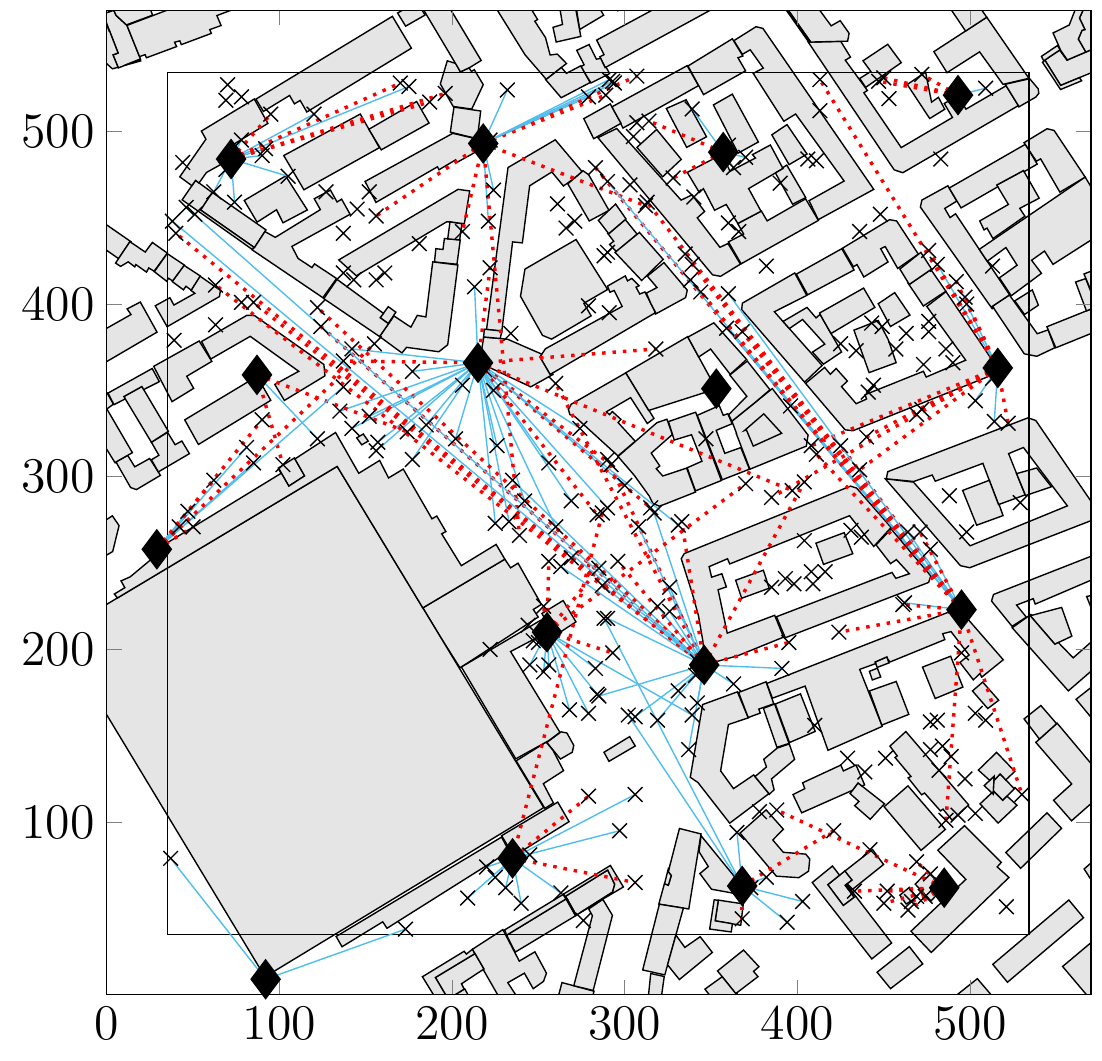}
	\caption{ \textit{SDMA, allocated links, \\ LOS ($\textcolor{cyan}{\rule[0.5ex]{0.4cm}{0.48pt}}$) and NLOS ($\textcolor{red}{--}$)}}
	 \label{fig:Frankfurt_heatmaps_i}
    \end{subfigure}
	\caption{Coverage heatmaps (SINR, UE throughput and link allocation) showing the effect of MAC schemes in Frankfurt ($\lambda_{BS}=64$~$\text{BSs/km}^2$, $\lambda_{UE}=1000$~$\text{UEs/km}^2$, $10^{\circ}$ ideal antennas).}
  	  \label{fig:Frankfurt_heatmaps}
\end{figure*}

	Let us first compare the performance of TDMA and SDMA for a nominal, baseline network configuration of $\lambda_{BS}=~64$~BSs/km$^2$,~$\lambda_{UE}=~1000$~UEs/km$^2$ and $10^{\circ}$ ideal antennas. Figs.~\ref{fig:1a} and \ref{fig:1b} present the resulting SINR and UE throughput, respectively, in Frankfurt and Seoul. We also show the interference-free SU case as an upper-bound reference. Fig.~\ref{fig:3} presents the corresponding interference component distributions, and Fig.~\ref{fig:Frankfurt_heatmaps} illustrates the SINR, UE throughput and allocated LOS/NLOS links for an example network realization in Frankfurt. 

	Fig.~\ref{fig:1a} shows that the deployed network reaches a coverage ratio of $70\%$ in Frankfurt, where the dense building layout causes substantial shadowing thus limiting coverage. The SINR distribution for TDMA is close to the SU bound, with a median SINR of around $30$~dB and $35$~dB respectively. The small gap between the two, which can also be observed in the SINR heatmaps in Figs.~\ref{fig:Frankfurt_heatmaps_a}~and~\ref{fig:Frankfurt_heatmaps_b}, is chiefly due to the link allocation heuristic assigning different links (along LOS/NLOS propagation paths) to some UEs for TDMA vs. SU, e.g. several cell-edge UEs around $(150$~m,~$350$~m$)$ in Figs.~\ref{fig:Frankfurt_heatmaps_g}~and~\ref{fig:Frankfurt_heatmaps_h}. This also demonstrates that our heuristic effectively mitigates the inter-cell interference that would otherwise result in a multi-user network were the UEs assigned their RSS-maximizing link as in SU. By contrast, Fig.~\ref{fig:1a} shows that the SDMA SINR distribution is significantly left-shifted since UEs suffer from inter-cell \emph{and} intra-cell interference, which also results in different links being allocated for SDMA. In particular, Fig.~\ref{fig:Frankfurt_heatmaps_i} shows that the SDMA link allocation consists of a large number of NLOS links, as the heuristic favors mitigating intra-cell interference over allocating the best per-user link with the highest SNR. We note that the jump in the SDMA SINR distributions around $0$~dB in Fig.~\ref{fig:1a} is an artifact of the ideal sectored antenna pattern\footnote{Nearby UEs, at a similar distance to the BS and served by beams with an orientation within the HPBW, experience strong intra-cell interference ($S~\approx~I$) since the received and interfereing signals are amplified by the same main-lobe gain as per the sectored ideal antenna model. Similarly, we observe a smaller jump around $-3$~dB, corresponding to the less likely case of intra-cell interference from two neighbor UEs within the HPBW ($S~\approx~I/2$).} (\emph{cf.} results for realistic non-discrete antenna patterns in Sec.~\ref{sec:ResAntennas}). Fig.~\ref{fig:1a} also shows that the considered network deployment is able to achieve full coverage in the open-spaced Seoul, where we observe the same trends in terms of the SINR distribution, with a median SINR of around $25$~dB for SDMA and $39$~dB for TDMA and SU.  
	
	The overall SINR performance of the two MAC schemes can be explained by examining the interference component distributions in Fig.~\ref{fig:3}, where we make two important observations. Firstly, the inter-cell interference is mostly below the noise floor ($-78$~dBm), making its effect negligible for all but a few UE cases in both SDMA and TDMA. We note that the inter-cell interference is slightly lower for TDMA than for SDMA due to the air-time sharing among interfering links modeled as $a_r$ in (\ref{equ:inter}). Secondly, Fig.~\ref{fig:3} shows that significant (above noise floor) intra-cell interference affects around 25\% and 20\% of all UEs in Frankfurt and Seoul, respectively. This demonstrates that intra-cell interference is overwhelmingly the limiting factor for SDMA performance.	

Nonetheless, Fig.~\ref{fig:1b} shows that SDMA significant	ly outperforms TDMA in terms of UE throughput for both study areas. SDMA's ability to simultaneously serve multiple UEs results in higher throughput even when taking into account the increased aggregate (intra-cell and inter-cell) interference and correspondingly lower SINR compared to TDMA. By contrast, air-time sharing limits the throughput performance of TDMA, despite better SINR conditions, as also illustrated in Fig.~\ref{fig:Frankfurt_heatmaps}.

\begin{figure}[tb!]
	\begin{subfigure}[]{0.5\textwidth}
  	 \includegraphics[width=0.95\columnwidth]{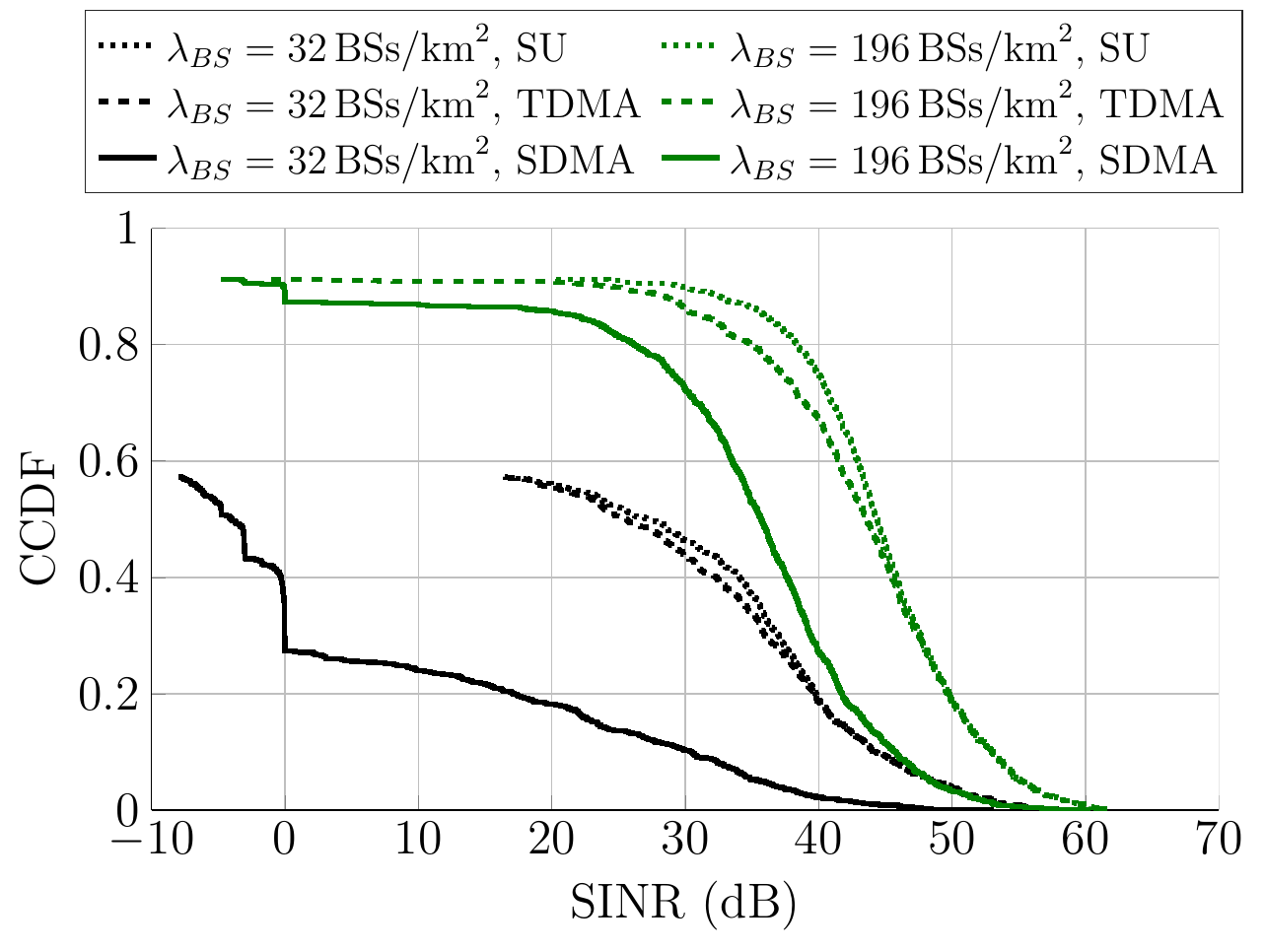}
  	 \caption{\textit{Frankfurt}}
  	    	 \label{fig:8a}
	\end{subfigure}
	~
	\begin{subfigure}[]{0.5\textwidth}
  	 \includegraphics[width=0.95\columnwidth]{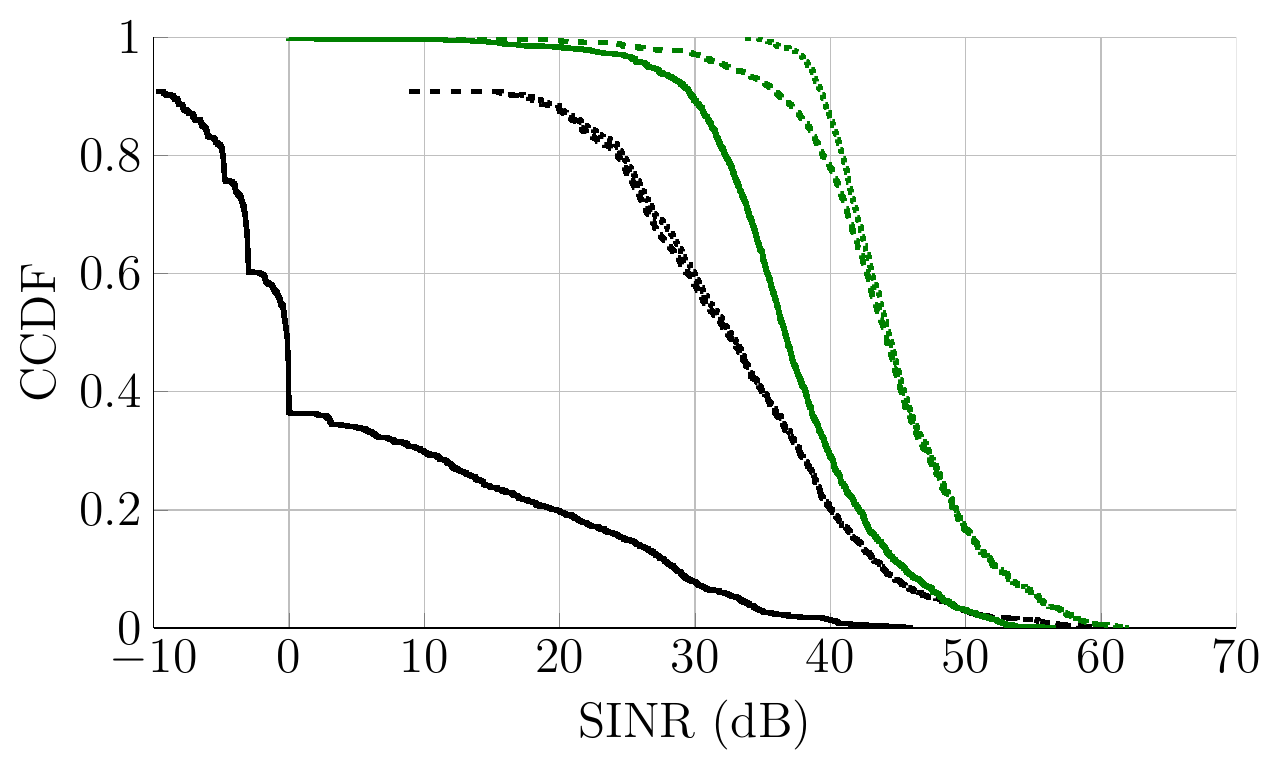}
  	 \caption{\textit{Seoul}}
  	    	 \label{fig:8b}
	\end{subfigure}
   	 \caption{SINR distributions showing the effect of BS density ($\lambda_{UE}=1000$~$\text{UEs/km}^2$, $10^{\circ}$ ideal antennas).}
   	 \label{fig:8}
\end{figure}

\begin{figure}[tb!]
	\begin{subfigure}[]{0.5\textwidth}
	\includegraphics[width=0.95\columnwidth]{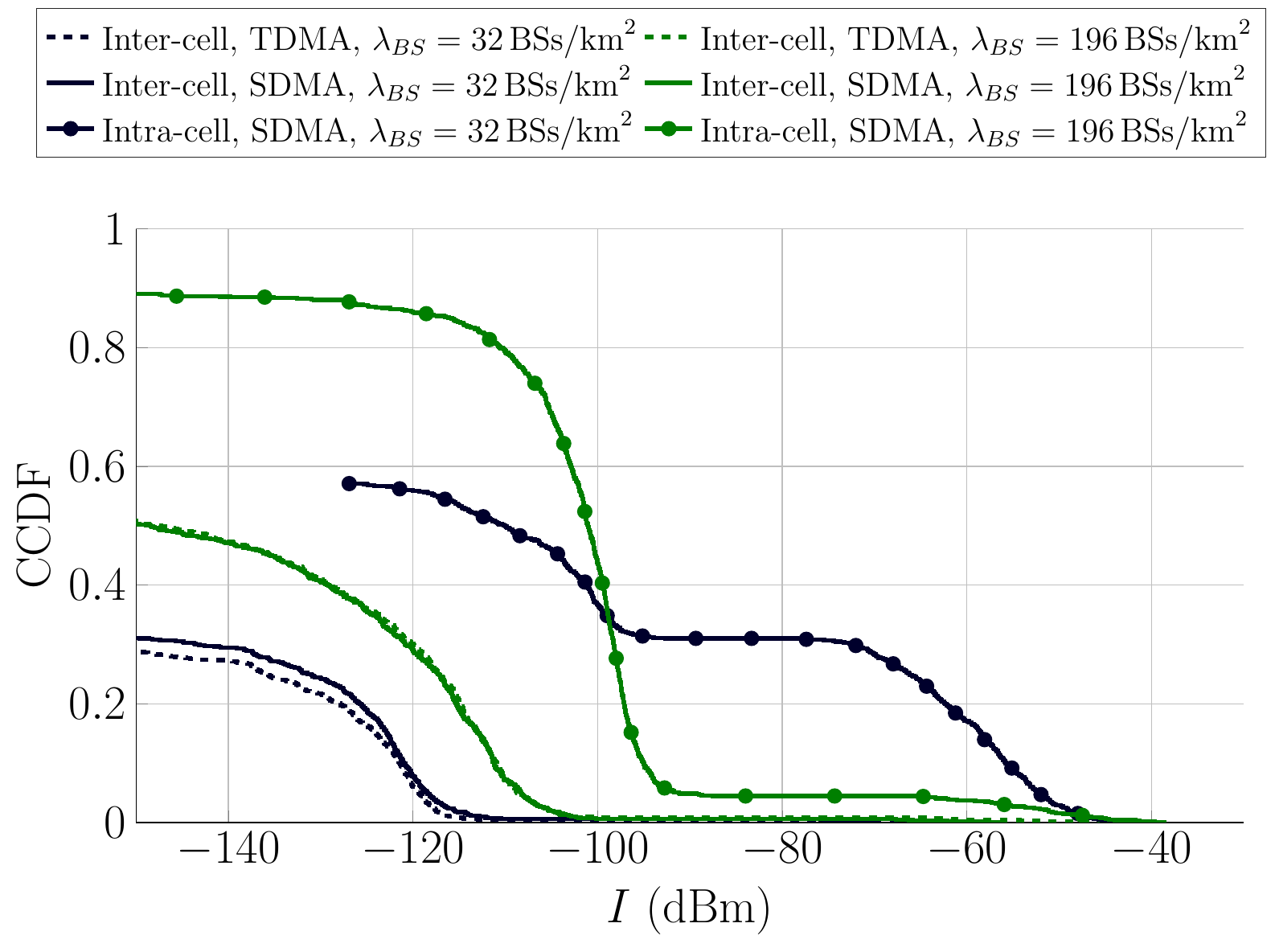}
	\caption{ \textit{Frankfurt}}
	   	 \label{fig:9a}
   	 \end{subfigure}
	~
	\begin{subfigure}[]{0.5\textwidth}
	\includegraphics[width=0.95\columnwidth]{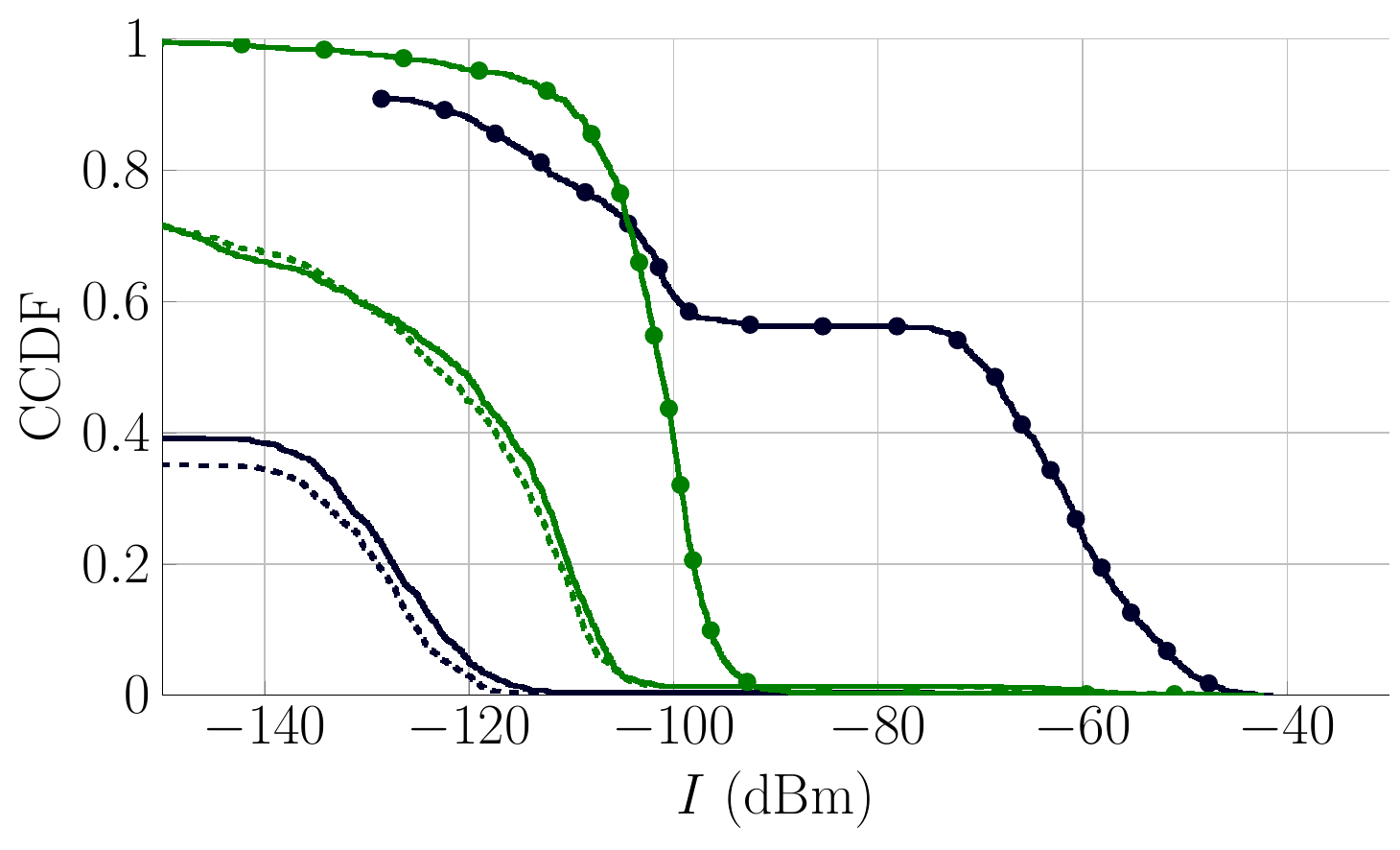}
	\caption{ \textit{Seoul}}
	   	 \label{fig:9b}
   	 \end{subfigure}
   	 \caption{Interference component distributions showing the effect of BS density ($\lambda_{UE}=1000~\text{UEs/km}^2$, $10^{\circ}$ ideal antennas).}
   	 \label{fig:9}
\end{figure}

\subsection{Effect of the Network Density}
\label{sec:deployment}

	First, we analyze the effect of BS density by considering two deployments, a sparse one of $\lambda_{BS}=32~\text{BSs/km}^2$ and a dense one of $\lambda_{BS}=196~\text{BSs/km}^2$. Fig.~\ref{fig:8} shows the corresponding SINR distributions for different MAC schemes in Frankfurt and Seoul, assuming $\lambda_{UE}=1000$~$\text{UEs/km}^2$ and $10^{\circ}$ ideal antennas. The building-crowded layout in Frankfurt causes substantial shadowing and results in limited coverage of less than 60\% for the sparse deployment. The dense deployment results in a much higher coverage ratio of over $90$\%. The more open-space layout in Seoul results in a high coverage ratio of over $90$\% even for the sparse deployment. The six-fold increase of the BS density results in a median SINR gain of over $35$~dB and $15$~dB for SDMA and TDMA, respectively, for both study areas. These SINR gains from network densification are due to higher RSS (shorter link distances, increased LOS availability), as well as lower intra-cell interference for SDMA. The corresponding interference component distributions in Fig. ~\ref{fig:9} show that inter-cell interference does increase with densification due to shorter interfering link distances,  but remains negligible for both SDMA and TDMA. Importantly, Fig. ~\ref{fig:9} shows that BS densification very significantly reduces the ratio of UEs that experience significant (above noise floor) intra-cell interference for SDMA, from around $30$\% and $60$\% in sparse Frankfurt and Seoul deployments respectively, to less than $5$\% in the dense case. This is due to the smaller number of served UEs per BS and shorter link distances, which improve the BS's ability to efficiently spatially distinguish the associated UEs by pointing the individual beams, thus decreasing the probability of overlapping (interfering) beams. The densification gain in SDMA SINR is more pronounced in Seoul due to this site's better coverage and thus larger average number of \emph{servable} UEs per BS versus Frankfurt.
	
\begin{figure}[tb!]
    \centering
	\begin{subfigure}[]{0.5\textwidth}
	\includegraphics[width=0.95\columnwidth]{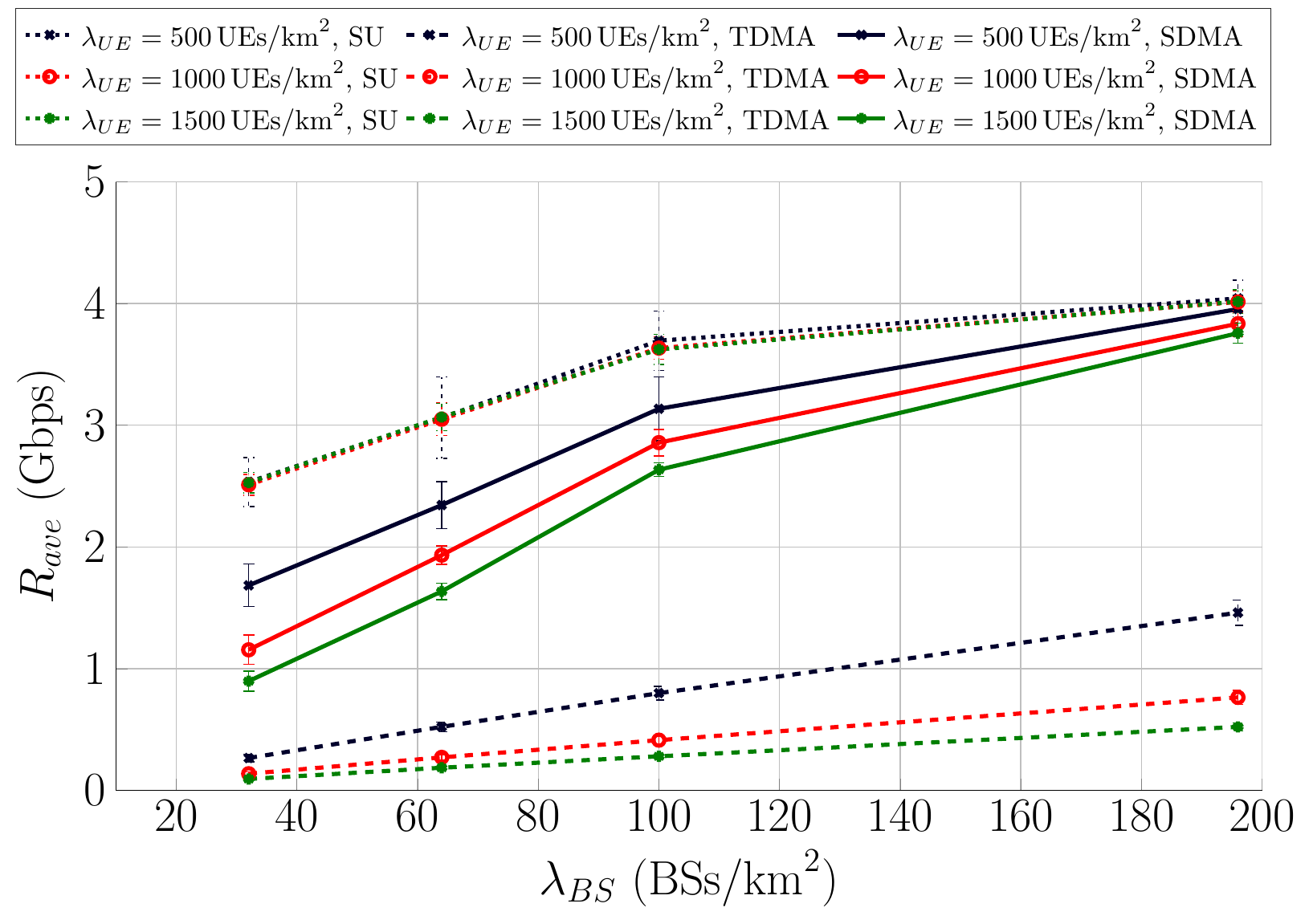}
	\caption{\textit{Frankfurt}}
	   	 \label{fig:10a}
    \end{subfigure}%
     
	\begin{subfigure}[]{0.5\textwidth}
	\includegraphics[width=0.95\columnwidth]{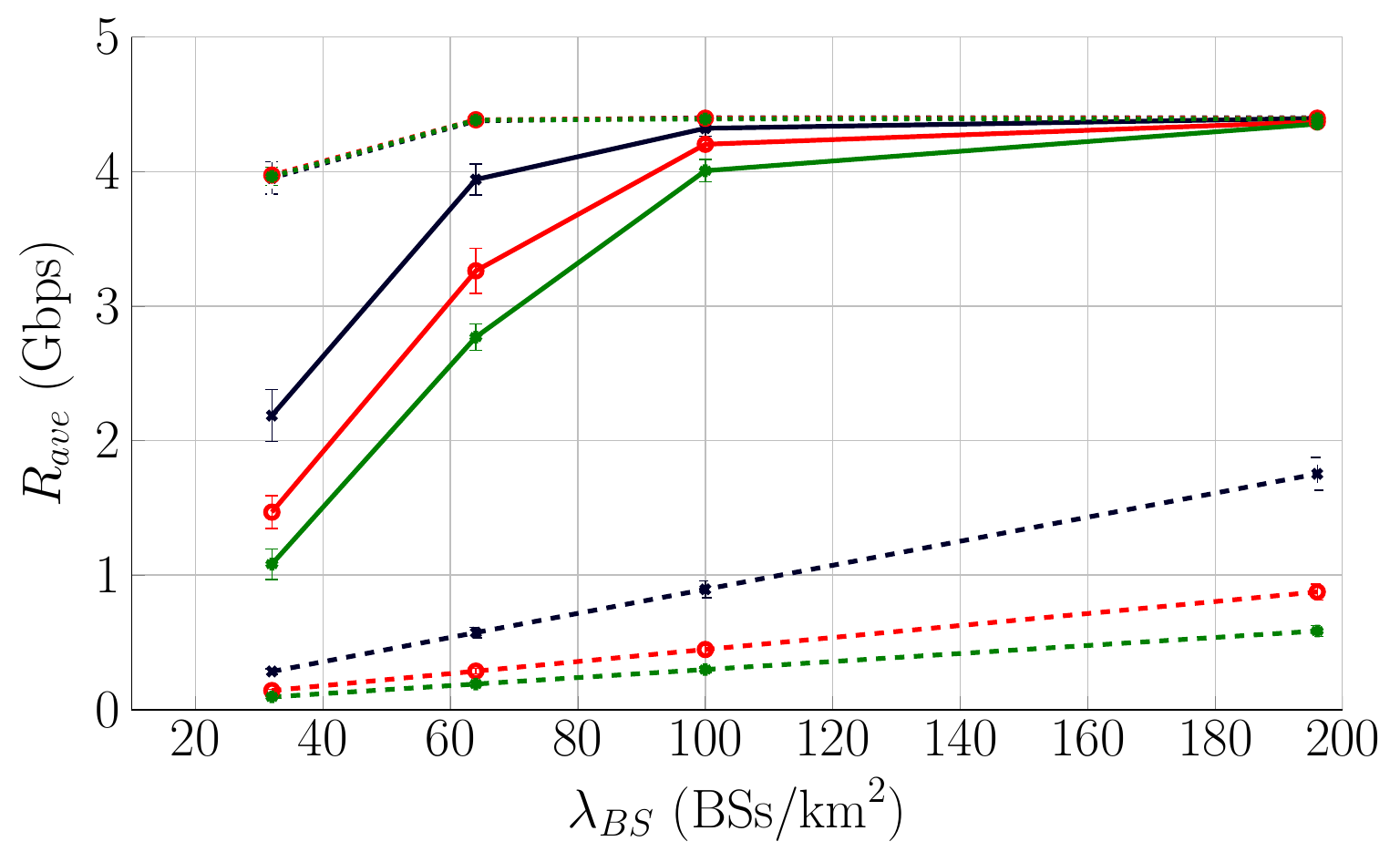}
	\caption{ \textit{Seoul}}
	   	 \label{fig:10b}
    \end{subfigure}
	\caption{Average UE throughput vs. BS density for different MAC schemes and UE densities ($10^{\circ}$ ideal antennas).}
  	  \label{fig:10}
\end{figure}

	Next, we consider in Fig.~\ref{fig:10} the average UE throughput versus the BS density for the different MAC schemes and different UE densities. Fig.~\ref{fig:10} confirms that the average UE throughput improves with the BS network densification and that this effect is more pronounced for SDMA than TDMA. The average UE throughput for SDMA in Seoul gradually saturates and performs very close to the upper SU bound for high BS densities over 100~$\text{BSs/km}^2$, as seen in Fig.~\ref{fig:10b}. This results from Seoul achieving almost full network coverage for $\lambda_{BS}=64~\text{BSs/km}^2$ (\emph{cf.}~Fig.~\ref{fig:1b}). This is not the case for Frankfurt, where even the highest BS density of $\lambda_{BS}=196~\text{BSs/km}^2$ does not reach full coverage (\emph{cf.}~Fig.~\ref{fig:1a}), which, in turn, results in the fact that even the SU bound does not achieve the maximum UE throughput of the considered data rate model in (\ref{equ:LTEthroughput}). This shows that there is still room for further SINR gains by deploying additional BSs in the coverage-challenged Frankfurt area, as there still exists a ratio of UEs that do not achieve the maximum UE throughput. The average UE throughput for TDMA also increases with BS densification, as a result of the smaller average number of served UEs per BS and correspondingly higher $a_r$.  	
		
	Let us now consider the average UE throughput with a given BS density for different UE densities, as shown in Fig.~\ref{fig:10}. As $\lambda_{UE}$ increases, Fig.~\ref{fig:10} shows a degradation in the average TDMA throughput consistent with the increase of the number of served UEs per BS, which dictates the air-time ratio -- the main limiting factor of TDMA performance. Consequently, the relative difference in TDMA throughput for different $\lambda_{UE}$ is roughly constant, but the absolute difference is more pronounced at higher $\lambda_{BS}$. The average SDMA throughput degradation as $\lambda_{UE}$ increases is likewise due to the larger number of UEs served per BS, which causes increased intra-cell interference -- the main limiting factor of SDMA performance. In contrast to TDMA, Fig.~\ref{fig:10} shows that the SDMA throughput discrepancy with different $\lambda_{UE}$ is less pronounced for higher $\lambda_{BS}$. This is because SDMA saturates towards the maximum UE throughput in dense networks, as discussed previously. Importantly, Fig.~\ref{fig:10} shows that for the sparse BS deployments of $\lambda_{BS}=32~\text{BSs/km}^2$, SDMA achieves an average throughput of around $1.1$~Gbps in Seoul, and $900$~Mbps in Frankfurt, for all UE densities. Fig.~\ref{fig:10} also shows that TDMA benefits consistently from BS densification, achieving over $500$~Mbps for the highest BS density of $\lambda_{BS}=196~\text{BSs/km}^2$, for all $\lambda_{UE}$. By contrast, at this highest network density, SDMA achieves an average throughput close to the SU bound, which is around $8$ times higher than TDMA. However, we expect the performance advantage of SDMA to be more limited if we remove the assumptions of ideal antenna beams and unlimited BS antenna resources, which we study in the sequel. For the remainder of our analysis, we present only the Frankfurt results for the sake of brevity, but we emphasize that the qualitative trends hold for both study areas.

\subsection{Effect of the Antenna Configuration} \label{sec:ResAntennas}

	We first consider in Sec.~\ref{sec:symAntennas} the case where the BS and the UEs are equipped with the same antennas  and study the effect of assuming ideal sectored antennas versus realistic antenna arrays with non-negligible sidelobes (\emph{cf}.~Sec.~\ref{sec:antennas}, Table~\ref{table:Antennas}). In Sec.~\ref{sec:asymAntennas}, we then consider the case where the BS is equipped with a higher gain antenna than the UE. We thereby investigate the achievable multi-user network performance when the UE has less stringent  beamforming requirements, which may be more attractive in practice, due to mobility management and device cost and complexity considerations. Throughout, we consider our baseline scenario of $\lambda_{BS}=64~\text{BSs/km}^2 $ and $\lambda_{UE}=1000~\text{UEs/km}^2$ in Frankfurt.

\subsubsection{Symmetric Antenna Configurations} \label{sec:symAntennas}

\begin{figure}[t]
    \centering
	\begin{subfigure}[]{0.5\textwidth}
	\includegraphics[width=0.95\columnwidth]{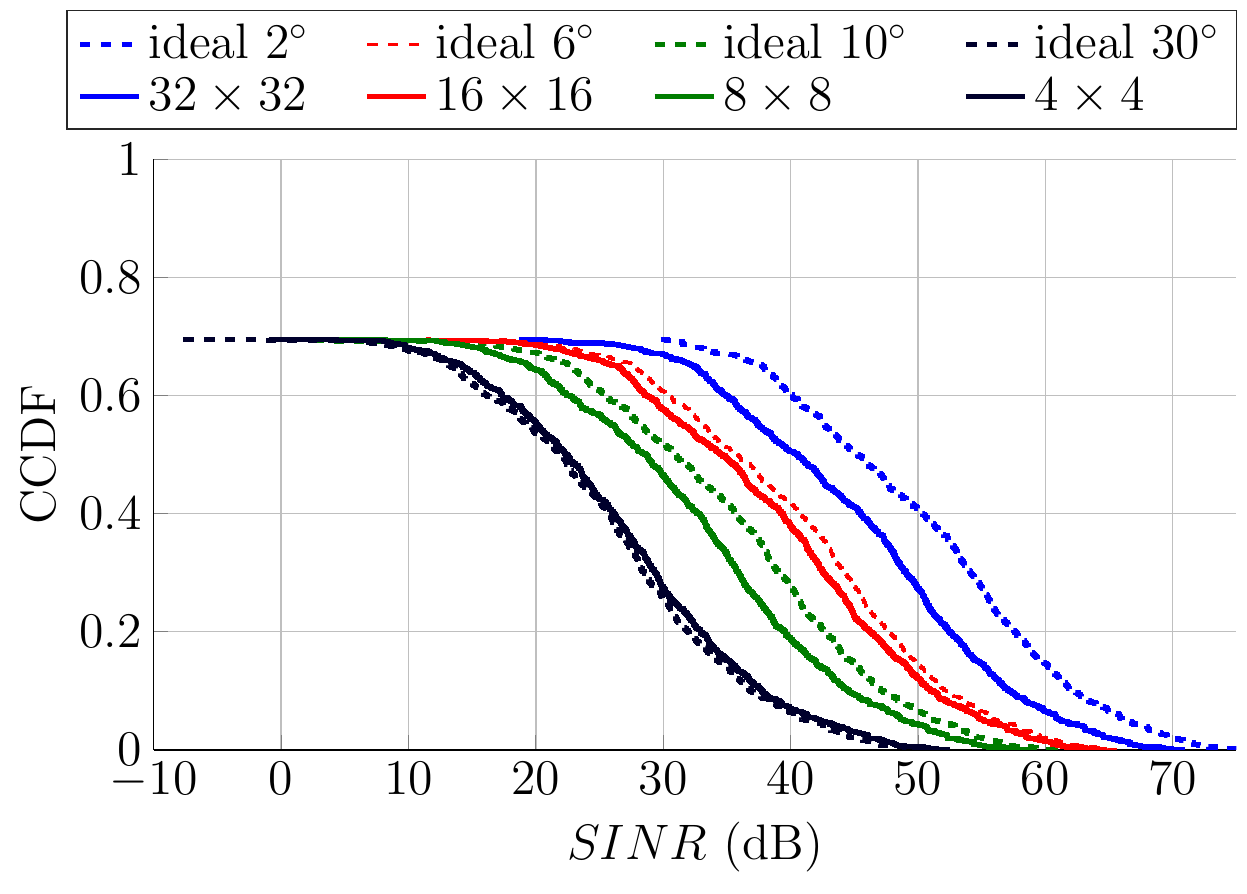}
	\caption{\textit{TDMA}}
	  	  \label{fig:13a}
    \end{subfigure}%
     
	\begin{subfigure}[]{0.5\textwidth}
	\includegraphics[width=0.95\columnwidth]{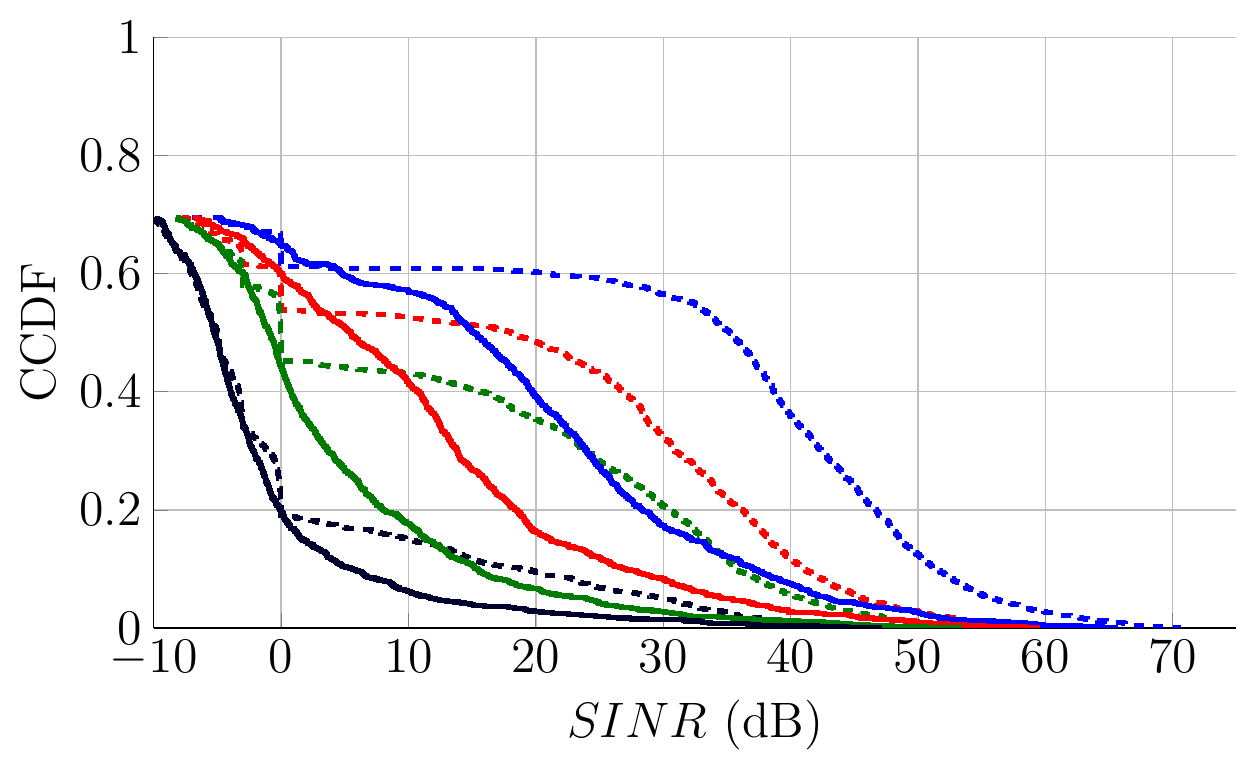}
	\caption{ \textit{SDMA}}
	  	  \label{fig:13b}
    \end{subfigure}
	\caption{SINR distributions for different MAC schemes and different symmetric BS/UE antenna configurations in Frankfurt ($\lambda_{BS}=64$~$\text{BSs/km}^2 $, $\lambda_{UE}=1000$~$\text{UEs/km}^2 $).}
  	  \label{fig:13}
\end{figure}

\begin{figure}[t]
    \centering
	\begin{subfigure}[]{0.5\textwidth}
	\includegraphics[width=0.95\columnwidth]{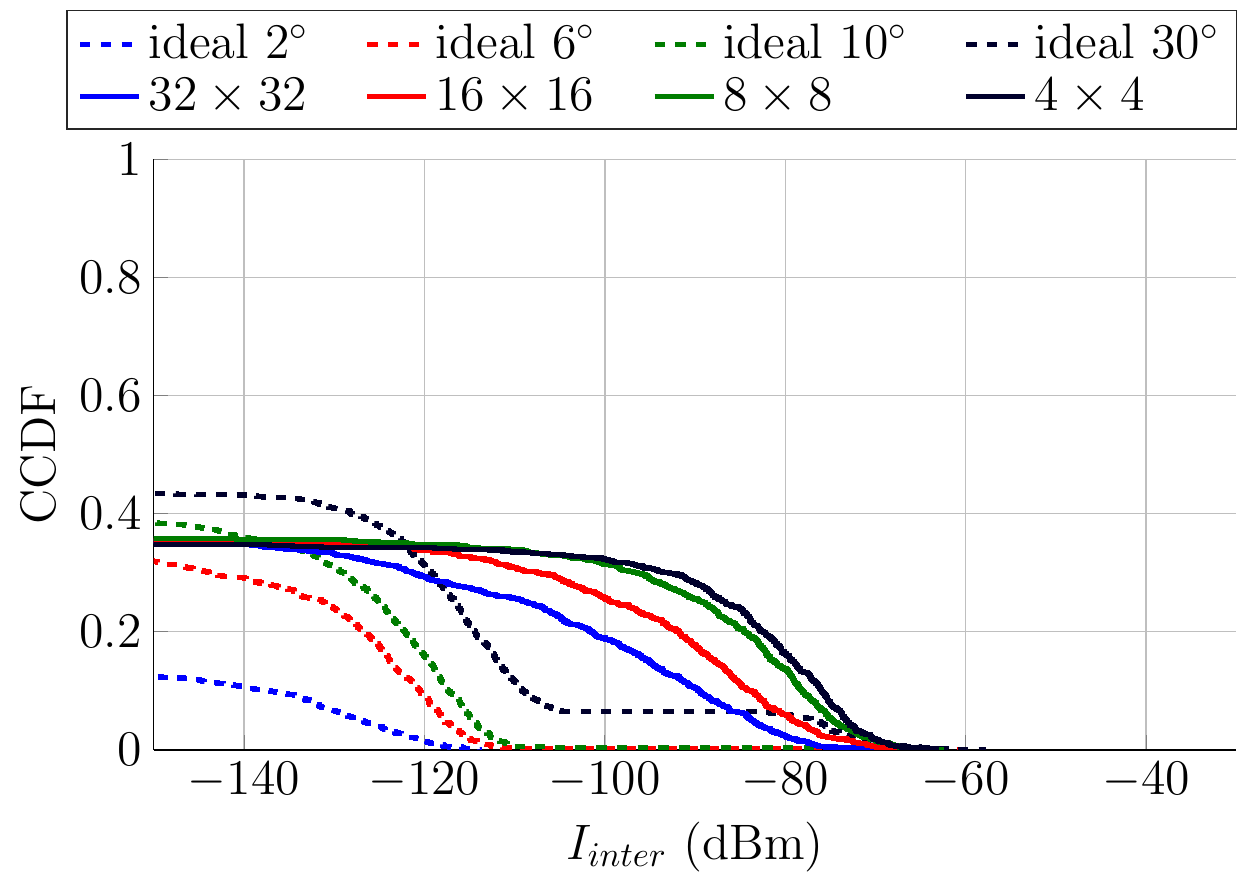}
	\caption{\textit{}}
	  	  \label{fig:14a}
    \end{subfigure}%
     
	\begin{subfigure}[]{0.5\textwidth}
	\includegraphics[width=0.95\columnwidth]{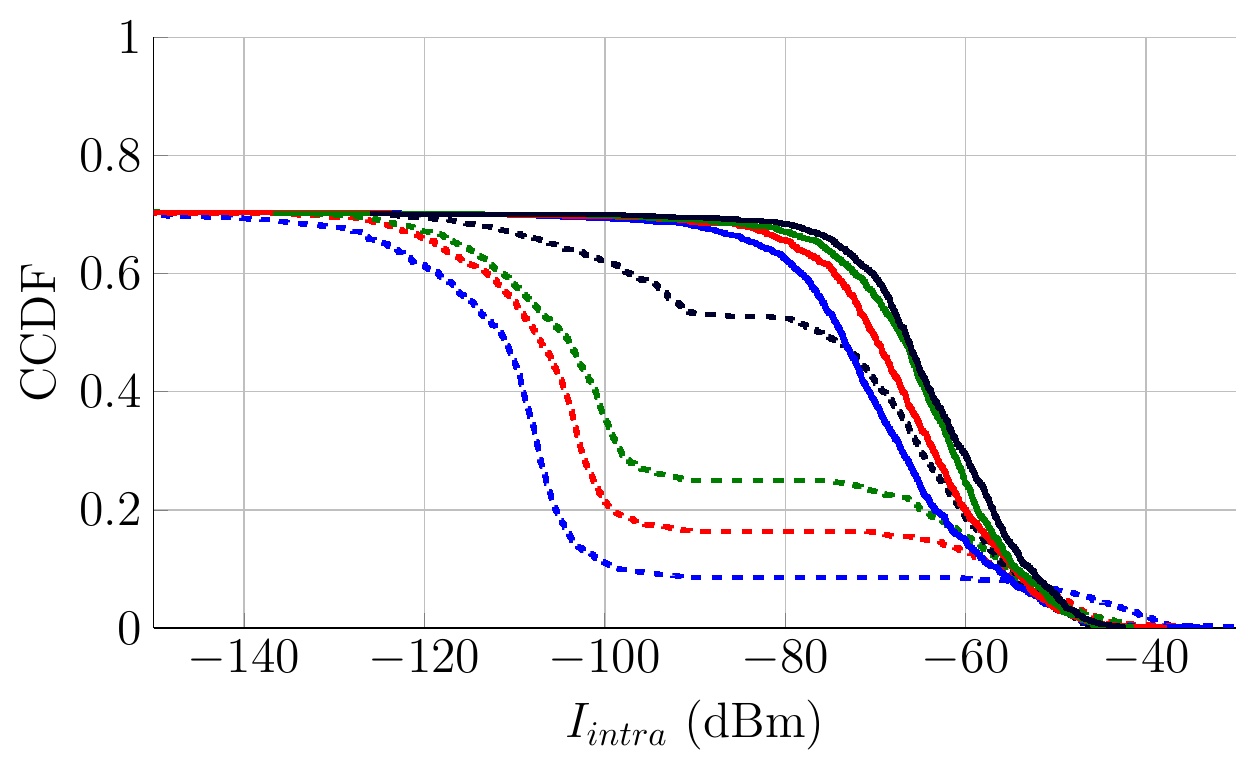}
	\caption{\textit{}}
	  	  \label{fig:14b}
    \end{subfigure}
	\caption{Interference component distributions for SDMA and different symmetric BS/UE antenna configurations in Frankfurt ($\lambda_{BS}=64$~$\text{BSs/km}^2 $, $\lambda_{UE}=1000$~$\text{UEs/km}^2 $).}
  	  \label{fig:14}
\end{figure}

	Figs.~\ref{fig:13a}-\ref{fig:13b} present the SINR distributions for different symmetric antenna configurations for TDMA and SDMA, respectively. Fig.~\ref{fig:14} presents the corresponding interference component distributions for SDMA. Fig.~\ref{fig:13} shows that the SINR increases with increasing antenna directionality, due to a higher RSS via higher main lobe gain and the smaller interference footprint of narrower beams. We also observe that SDMA is more sensitive to the antenna directionality than TDMA, regardless of antenna pattern type. For example, moving from a $10^{\circ}$ to a $6^{\circ}$ ideal antenna yields a median SINR gain of $5$~dB for TDMA and $20$~dB for SDMA. This is because narrower antenna beams result in a marked decrease in significant intra-cell interference for SDMA, as shown in Fig.~\ref{fig:14b}. By contrast, Fig.~\ref{fig:14a} shows that inter-cell interference is mostly negligible relative to the noise floor, even for wide-beam ideal antenna (and TDMA inter-cell interference is even slightly lower than for SDMA, \emph{cf.} Fig.~\ref{fig:3}).
	
	Importantly, Fig.~\ref{fig:13} shows that considering a realistic model of antenna arrays instead of ideal sectored antennas results in a degradation of SINR for patterns with comparable main lobe gains. For SDMA, this is a result of the very significant increase in strong (above noise floor) intra-cell interference observed in Fig.~\ref{fig:14b}, due to the non-negligible sidelobes in the realistic antenna arrays and the resulting increase in spatial overlap among per-UE beams. Consequently, realistic modeling of sidelobes for SDMA results in a gap of up to $20$~dB compared to the SINR distribution assuming ideal sectored antennas\footnote{Fig.~\ref{fig:13b} confirms that the jumps in the SDMA distribution for ideal antennas are an artifact of the discrete sectored antenna model.}  in Fig.~\ref{fig:13b}. Fig.~\ref{fig:13b} also shows that the effect of modeling sidelobes -- and the resulting interference -- is more prominent for highly directional antennas, due to a lower likelihood of the main lobes overlapping for narrow beams, as reflected in the low intra-cell interference for narrow-beam ideal antennas in Fig.~\ref{fig:14b}. By contrast, the effect of a non-ideal antenna pattern is rather minor for TDMA, since sidelobe modeling results in only a minor increase in above-noise-floor inter-cell interference, as evident in Fig.~\ref{fig:14a}; instead, the TDMA SINR degradation observed for realistic vs. ideal antennas, \emph{cf.} Fig.~\ref{fig:13a} is mostly due to a slight mismatch in the main lobe gain of the two antenna types (e.g. $2.2$~dBi higher for the $8\times8$ array than the $10^{\circ}$ ideal antenna in Table~\ref{table:Antennas}). 
		
	\begin{figure}[h!]
	\centering
	\includegraphics[width=0.95\columnwidth]{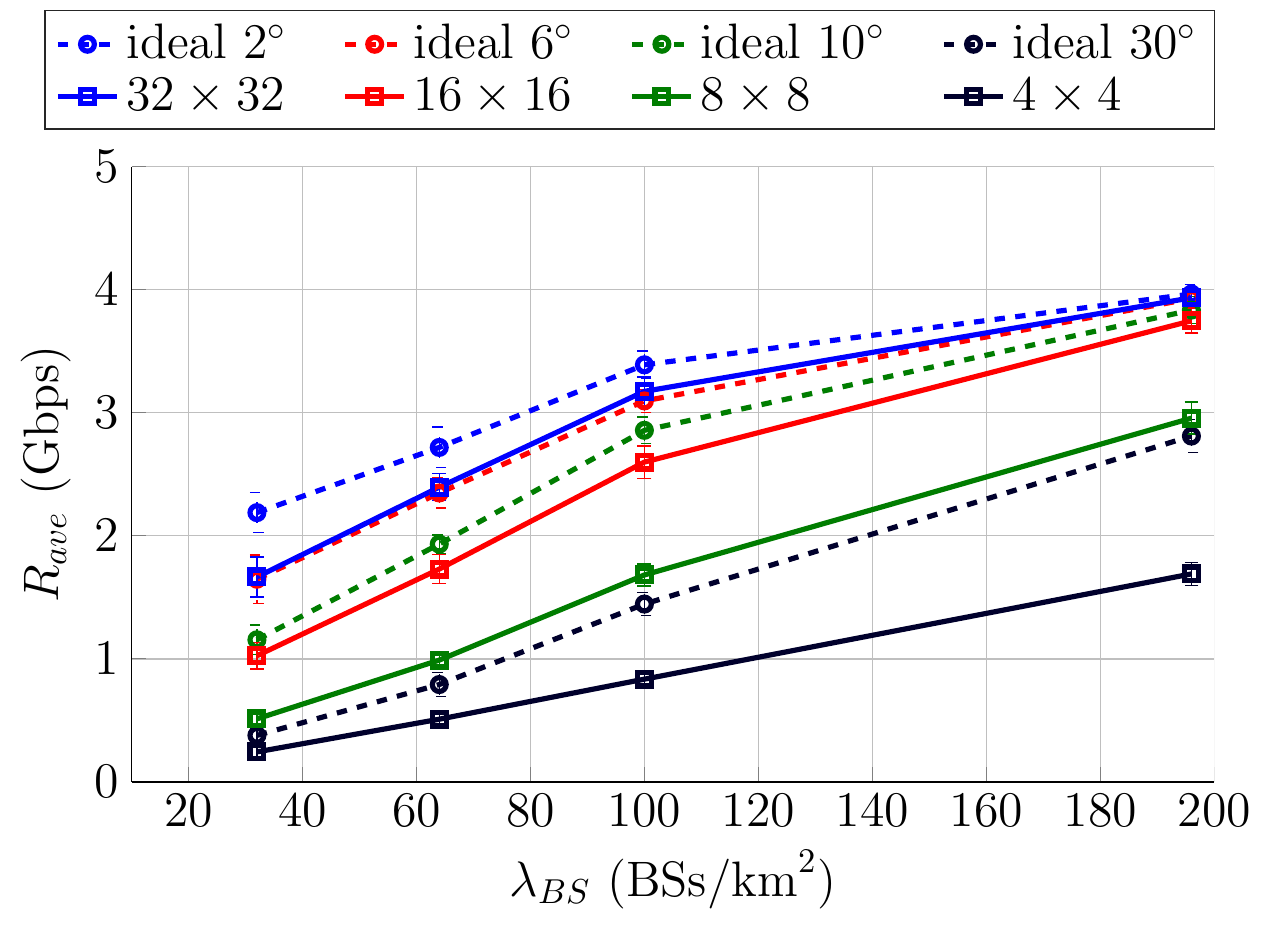}
	\caption{Average UE throughput for SDMA vs. BS density, showing the effect of different symmetric BS/UE antenna configurations in Frankfurt ($\lambda_{UE}~=~ 1000\,\text{UEs/km}^2$).}
    	\label{fig:15}
\end{figure}

	The importance of realistic antenna modeling in estimating SDMA performance is evident in the average UE throughput shown in Fig.~\ref{fig:15} versus BS density, for different symmetric antenna configurations. For example, assuming $8\times8$ arrays rather than ideal $10^{\circ}$ ideal antennas for our baseline 	$\lambda_{BS}=64$~$\text{BSs/km}^2$ results in a decrease of around $50\%$, i.e. $1$~Gbps, in the average UE throughput. However, Fig.~\ref{fig:15} also shows that the discrepancy between the ideal and realistic antenna models decreases for high BS densities, due to SDMA throughput saturating towards the maximum in dense networks when the SINR exceeds $SINR_{max}$ (as discussed for Fig.~\ref{fig:10}). The same effect is evident in Fig.~\ref{fig:15} for more highly directional antennas, where the discrepancy in SINR estimated using realistic versus ideal antennas is less prominent in the resulting throughput. Finally, we note that we omit for brevity the average UE throughput results for TDMA, since TDMA average throughput is largely \emph{insensitive} to both antenna model and directionality. Namely, TDMA throughput results for all antenna configurations in Fig.~\ref{fig:10} are nearly identical to the $\lambda_{UE}=1000$~$\text{BSs/km}^2$ TDMA curve in Fig.~\ref{fig:10a}. This follows simply from the TDMA SINR distribution in Fig.~\ref{fig:13a} approaching or exceeding $SINR_{max}$ for all antenna configurations, so that the dominant limiting factor is the air-time sharing ratio among UEs served by a given BS, which is independent of the antenna configuration. Consequently, SDMA still outperforms TDMA in average UE throughput by up to $1$~Gbps with $8\times8$ arrays.

\subsubsection{Asymmetric Antenna Configurations} \label{sec:asymAntennas}

\begin{figure}[t]
    \centering
	\begin{subfigure}[]{0.5\textwidth}
	\includegraphics[width=0.95\columnwidth]{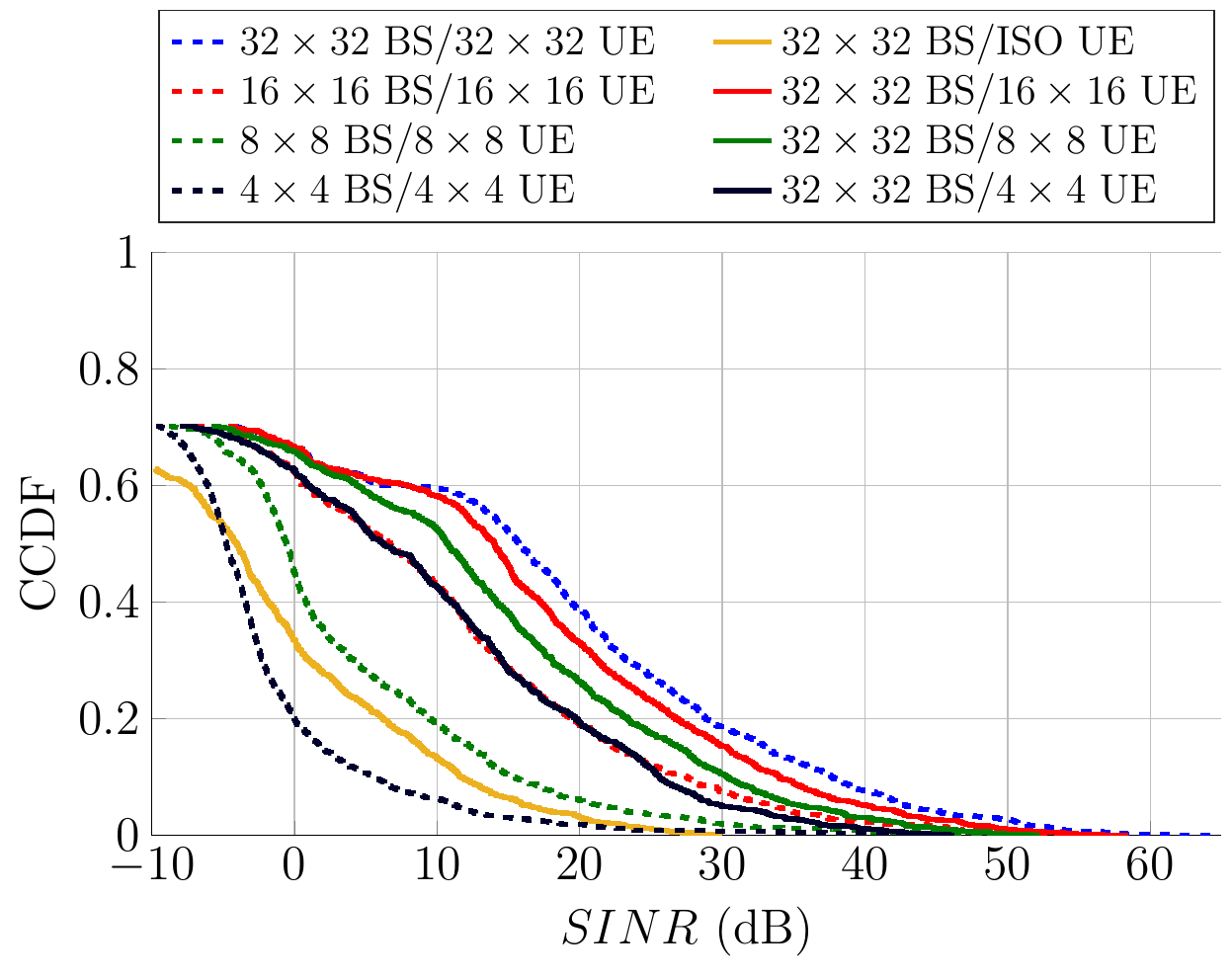}
	\caption{\textit{}}
	  	 \label{fig:16a}
    \end{subfigure}
	\begin{subfigure}[]{0.5\textwidth}
	\includegraphics[width=0.95\columnwidth]{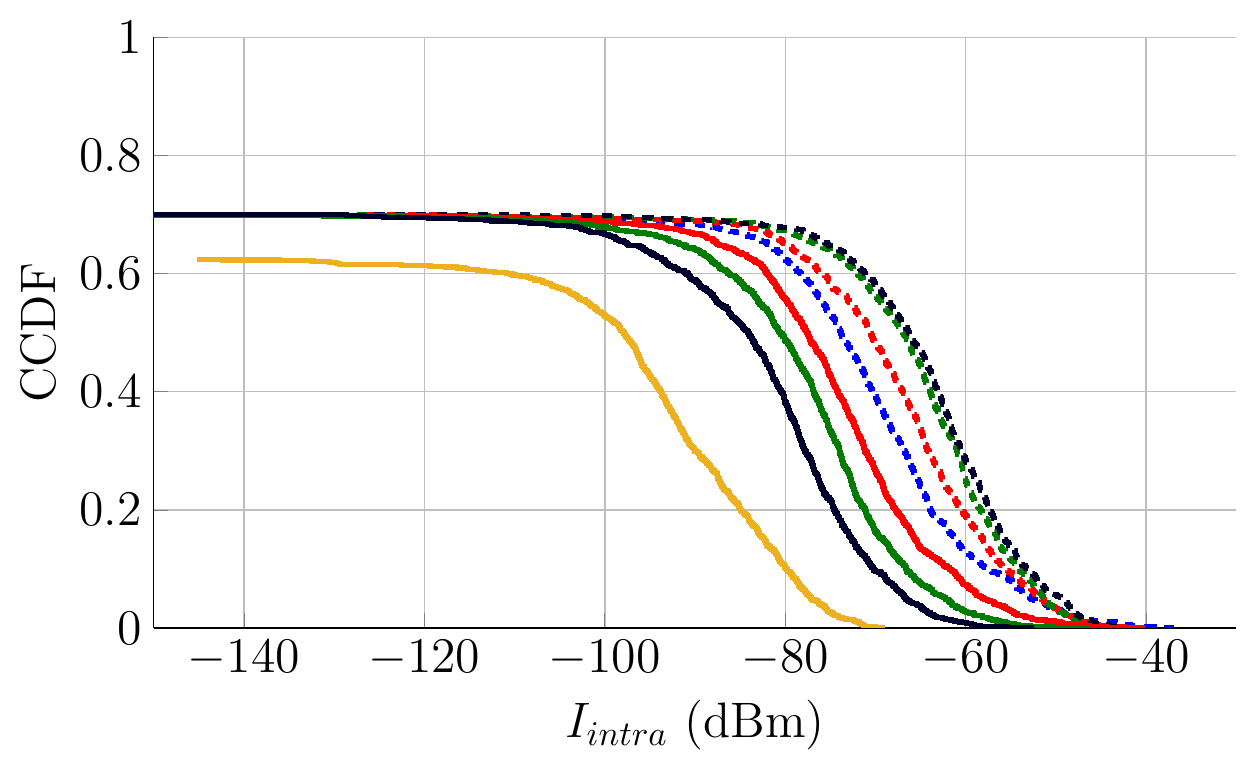}
	\caption{ \textit{}}
	  	 \label{fig:16b}
    \end{subfigure}
    	\caption{SINR and interference distributions for SDMA, using different asymmetric BS/UE antenna configurations in Frankfurt ($\lambda_{BS}=64$~$\text{BSs/km}^2 $, $\lambda_{UE}=1000$~$\text{UEs/km}^2 $).}
  	 \label{fig:16}
\end{figure}

\begin{figure}[h!]
	\centering
	\includegraphics[width=0.95\columnwidth]{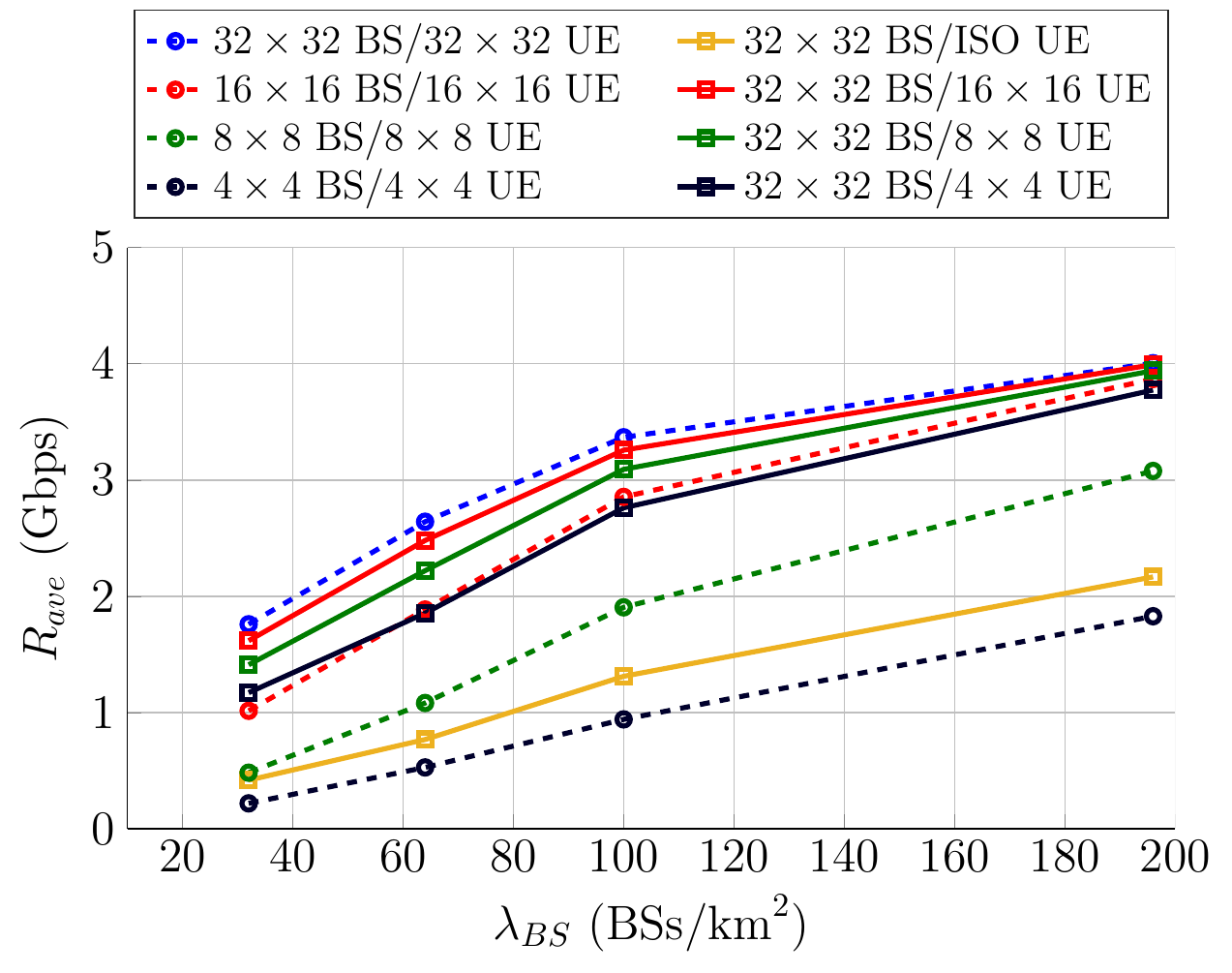}
	\caption{Average UE throughput for SDMA vs. BS density, showing the effect of different asymmetric BS/UE antenna configurations in Frankfurt ($\lambda_{UE}=1000\,\text{UEs/km}^2$).}
	
    	\label{fig:18}
\end{figure}

	Fig.~\ref{fig:15} shows that SDMA performance benefits from high antenna directivity, with $32\times32$ arrays achieving an average throughput of over 1.5~Gbps and 4~Gbps in sparse and dense deployments respectively. However, implementation of such large arrays is challenging, especially at the UE terminal with strict size, power, and cost constraints. Therefore, it may be more attractive to equip the BS with larger antenna arrays to compensate for a smaller array at the UE. To this end, we study the performance tradeoffs for such asymmetric BS/UE antenna configurations. Fig.~\ref{fig:16} presents the SINR and intra-cell interference for SDMA with $32\times32$ sub-arrays per UE served at the BS and different array sizes at the UE (including the omni-directional ISO antenna case). We also plot the reference symmetric BS/UE configurations of $16\times16$, $8\times8$ and $4\times4$. Fig.~\ref{fig:18} shows the corresponding average UE throughput versus BS density. To aid our comparison of the configurations, Table~\ref{table:linkBudget} gives their nominal interference-free link budgets; we note that the $EIRP_{max}$ limit at the BS means that the link budget is effectively determined by the UE gain.

		\begin{table}
			\centering
			\normalsize
			\begin{tabular}{l l l l}
					\toprule
					\textbf{Configuration} & \textbf{BS} & \textbf{UE} & \textbf{Link budget (dBm)} \\ 
					\midrule
					{} & 32$\times$32 & ISO &  -48 \\  		
					{asymmetric} & 32$\times$32 & 4$\times$4  & -31.5 \\	
					{} & 32$\times$32 & 8$\times$8  & -25.2 \\ 	
					{} & 32$\times$32 & 16$\times$16  & -19.1  \\   
					\hline
					{} & 32$\times$32 & 32$\times$32  & -13   \Tstrut \\  	
					{symmetric} & 16$\times$16  & 16$\times$16 &  -19.1 \\ 
					{} & 8$\times$8 & 8$\times$8 & -25.2  \\ 	  
					{} & 4$\times$4 & 4$\times$4 & -31.5 \\ 	
					\bottomrule
				\end{tabular} 
			\caption{Nominal link budget for different antenna configurations (10~m LOS link,  $EIRP_{max}=40$~dBm).}
			\label{table:linkBudget}
		\end{table}

	Fig.~\ref{fig:16a} shows that $32\times32$~BS/$4\times4$~UE achieves almost identical SINR to the symmetric $16\times16$ configuration, despite the asymmetric configuration having a 12~dB lower nominal link budget than the symmetric case (\emph{cf}.~Table~\ref{table:linkBudget}). Similarly, the $32\times32$~BS/ISO~UE configuration achieves a comparable SINR to both the  $4\times4$ and $ 8\times8$ symmetric configurations (mostly outperforming the former and being within $3$~dB of the latter), despite a nominal link budget deficit of around $25$~dB. The strong SINR performance of the asymmetric antenna configurations is a result of the $32\times32$ sub-arrays at the BS producing narrower per-UE beams which significantly reduces intra-cell interference, offsetting the lower RSS due the UE's lower gain antenna. Specifically, Fig.~\ref{fig:16b} shows that the asymmetric $32\times32$~BS/ISO~UE and $32\times32$~BS/$4\times4$~UE configurations result in median intra-cell interference below the noise floor, whereas all symmetric configurations are well above the noise floor for most of the covered UEs. 
	 
	The average UE throughput in Fig.~\ref{fig:18} is consistent with the SINR trends observed in Fig.~\ref{fig:16a}, confirming that deploying larger antenna arrays at the BS benefits SDMA performance by spatially limiting the intra-cell interference, while compensating for smaller UE arrays. This is a particularly encouraging result for practical deployments, since reduced UE antenna complexity is beneficial not only from the cost/power perspective, but also significantly relaxes UE beamsteering requirements. Notably, Fig.~\ref{fig:18} shows that average data rates of $420$~Mbps and $2.2$~Gbps are achievable in sparse and dense networks, respectively, even if UEs were to use omnidirectional antennas, which would immensely simplify mobility management. Fig.~\ref{fig:18} also shows that in a dense network with $32\times32$ BS arrays, the UE throughput achieved with the $4\times4$ UE array is around $3.75$~Gbps -- an improvement of $1.5$~Gbps over an omnidirectional UE antenna and within $6\%$ of the maximum $4$~Gbps throughput with a $32\times32$ UE array. Therefore, we consider the $32\times32$ BS/$4\times4$ UE antenna configuration to provide the best tradeoff between UE throughput and antenna complexity.

\subsection{HBF with a Limited Number of BS Sub-Arrays}
\label{sec:resHBF}

\begin{figure}[t]
	\begin{subfigure}[]{0.5\textwidth}
  	 \includegraphics[width=0.95\columnwidth]{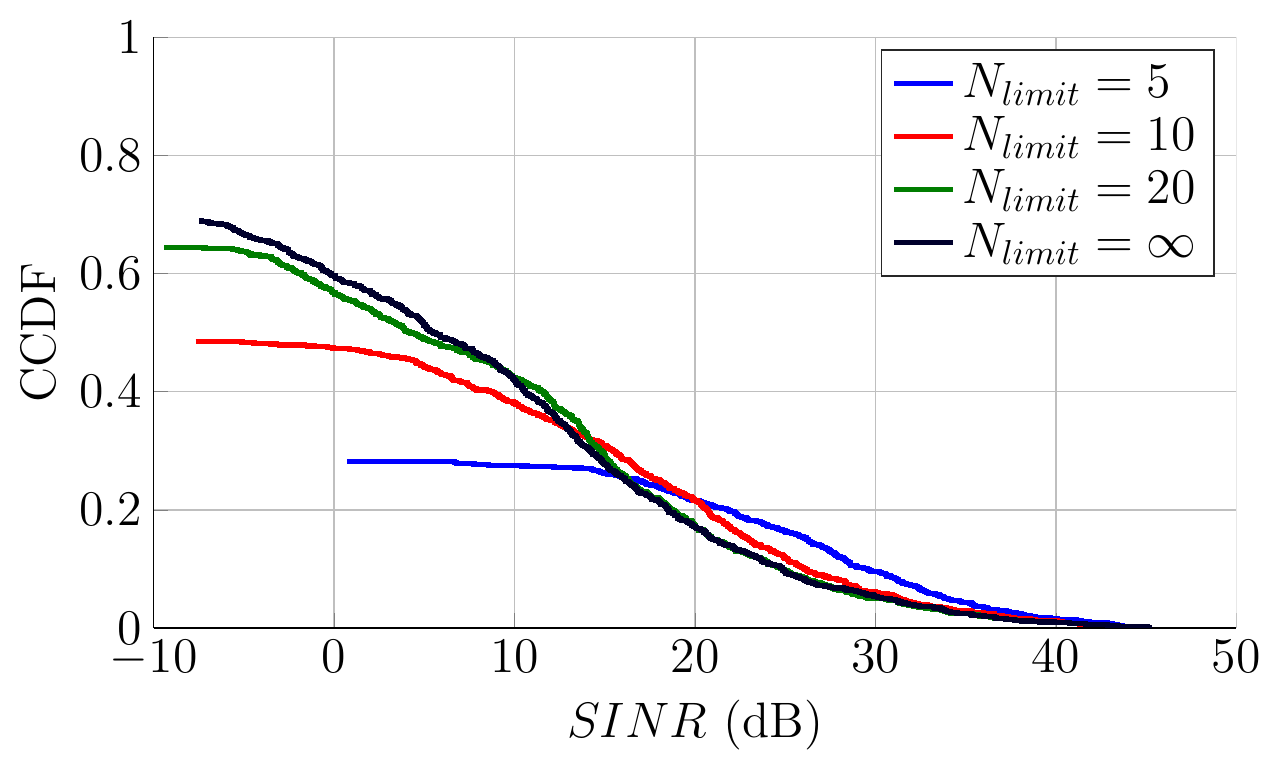}
  	 \end{subfigure}
	\caption{SINR distribution for SDMA with limited number of $32\times32$ BS sub-arrays in Frankfurt ($\lambda_{BS}=64$~$\text{BSs/km}^2$, $\lambda_{UE}=1000$~$\text{UEs/km}^2$, $4\times4$ UE array).}
  	 \label{fig:limitUEs_SINR_a}
\end{figure}

\begin{figure}[t]

	\begin{subfigure}[]{0.5\textwidth}
    \centering
	\includegraphics[width=0.95\columnwidth]{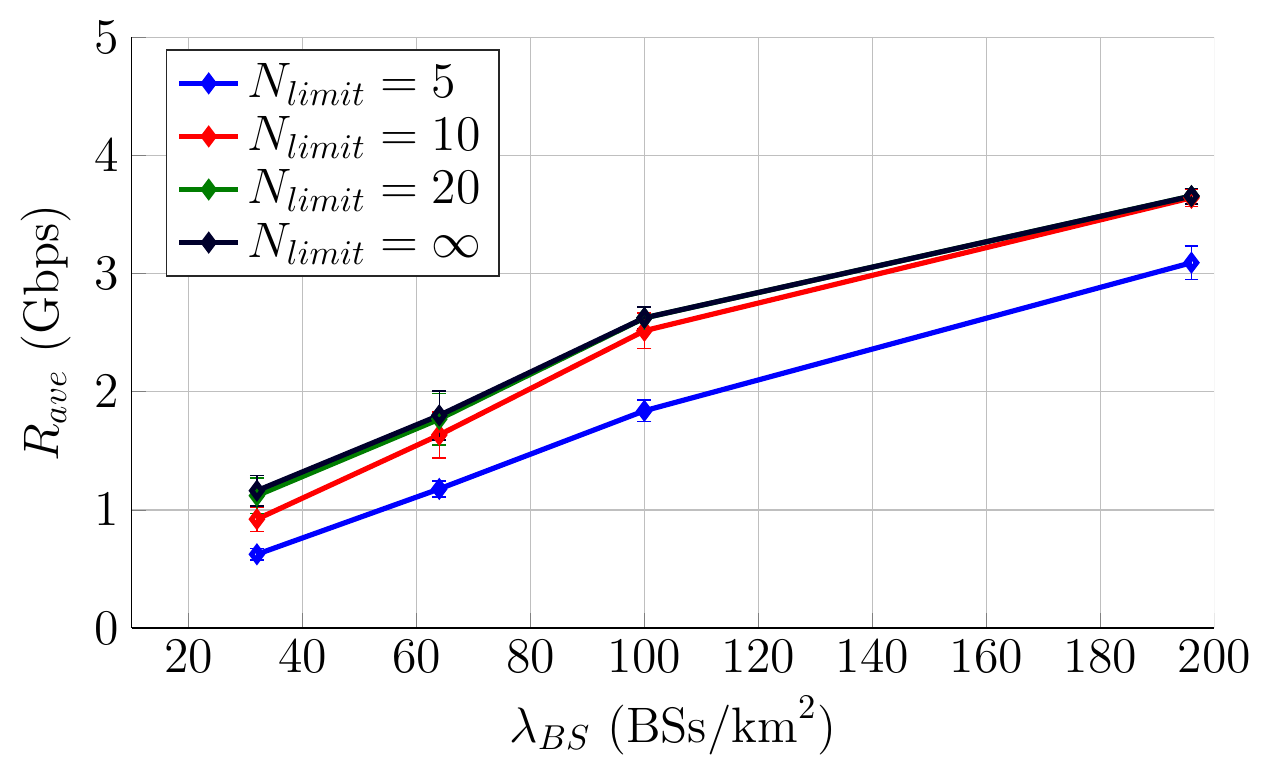}
	\caption{\textit{Average UE throughput}}
	 \label{fig:limitUEs_a}
    \end{subfigure}

	\begin{subfigure}[]{0.5\textwidth}
\centering
	\includegraphics[width=0.95\columnwidth]{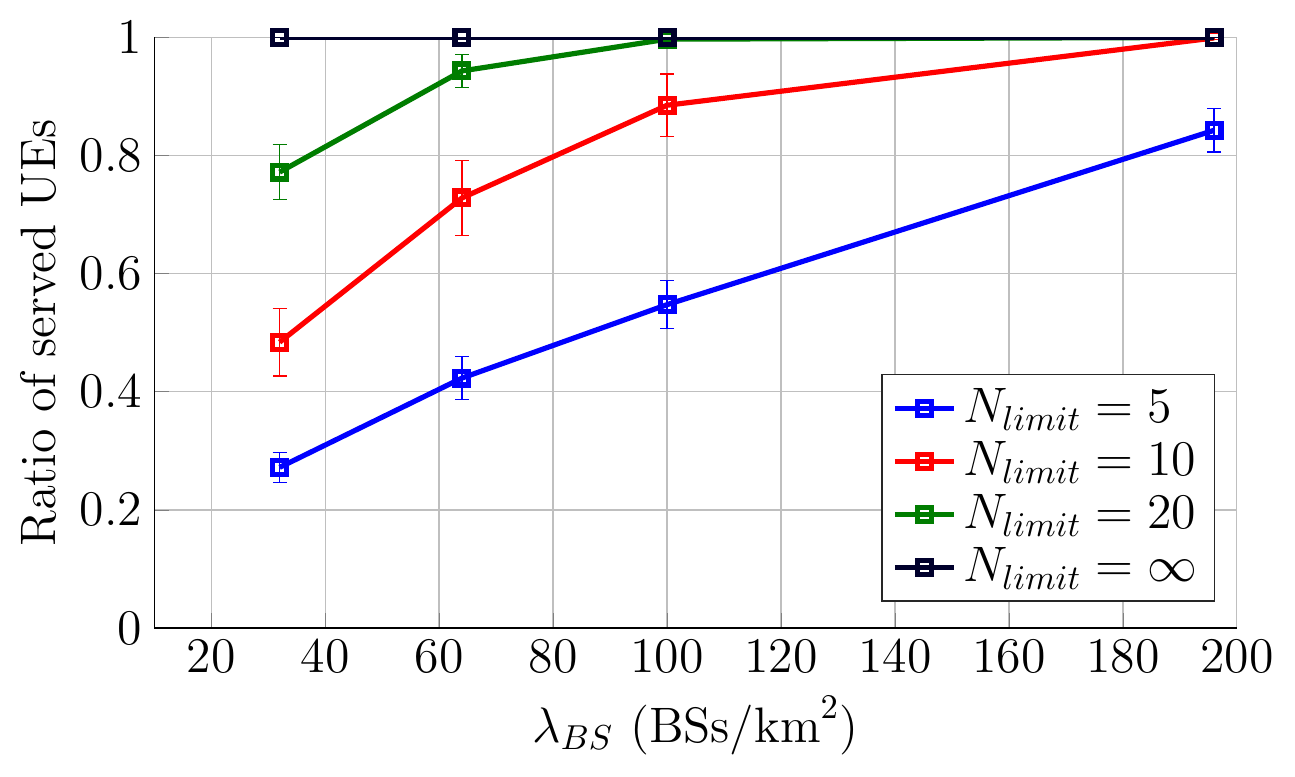}	
	\caption{\textit{Ratio of the served UEs (out of all servable UEs)}}
		 \label{fig:limitUEs_b}
    \end{subfigure}
	\caption{Average UE throughput and ratio of served UEs vs. BS density for SDMA with limited number of $32\times32$ BS sub-arrays in Frankfurt ($\lambda_{UE}=1000$~$\text{UEs/km}^2$, $4\times4$ UE array).}
	 \label{fig:limitUEs}
\end{figure}

	In our analysis thus far we have adopted the simplifying assumption for SDMA that the BS may simultaneously serve an unlimited number of UEs. Since in a practical HBF architecture each served UE corresponds to an additional antenna sub-array at the BS, we study the performance of SDMA with a limited number of BS sub-arrays, $N_{limit}=\{5, 10, 20\}$.
	
	Fig.~\ref{fig:limitUEs_SINR_a} presents the SINR distributions for SDMA with different $N_{limit}$ of $32\times32$ BS sub-arrays, with a $4\times4$ array at each UE and our baseline network scenario (Frankfurt, $\lambda_{BS}=64$~$\text{BSs/km}^2$, $\lambda_{UE}=1000$~$\text{UEs/km}^2$). Fig.~\ref{fig:limitUEs_SINR_a} shows that limiting the number of BS sub-arrays reduces the overall network coverage ratio due to dropped UEs, e.g. down from $70$\% for $N_{limit}=\infty$ to only $30$\% for $N_{limit}=5$. Fig.~\ref{fig:limitUEs_SINR_a} also shows that the non-dropped UEs generally achieve a slightly higher SINR with a smaller $N_{limit}$ (e.g. $N_{limit}=5$ gives the highest cell-edge SINR and the highest proportion of UEs achieving an SINR above $20$~dB).  This is a result of inherently reduced intra-cell interference\footnote{The interference distributions corresponding to Fig.~\ref{fig:limitUEs_SINR_a} show a straightforward reduction of $I_{intra}$ with decreasing $N_{limit}$ and are omitted for brevity.} for small $N_{limit}$, combined with our link allocation heuristic prioritizing serving UEs with the highest achievable performance (\emph{cf}. Sec.~\ref{sec:Link_allocation_heuristic}).

	Given UEs with good SNR coverage may be dropped due to BS antenna constraints, in Fig.~\ref{fig:limitUEs} we show the overall network performance of SDMA with $N_{limit}<\infty$ versus BS density in terms of the average UE throughput and the ratio of served UEs (out of all servable UEs, i.e. compared to SDMA with $N_{limit}=\infty$). Fig.~\ref{fig:limitUEs_a} shows that $N_{limit}=10$ is sufficient to achieve an average throughput comparable to the unlimited antenna resources, regardless of BS density. However, the $N_{limit}=10$ limit comes at the cost of up to 50\% of all servable UEs being dropped in sparser networks, as shown in Fig.~\ref{fig:limitUEs_b}. This result highlights the tradeoff between BS densification and antenna resource constraints -- for sparse deployments, the BS antenna resource requirements are higher due to the higher per-BS load of servable UEs, and thus, a limited number of BS sub-arrays results in a poorer coverage ratio. Conversely, with BS densification, the BS antenna resource requirements are eased due to the lower per-BS load of servable UEs.

\section{Conclusions}
\label{sec:Conclusions}

	We presented the first comprehensive study of the performance of TDMA and SDMA mm-wave urban cellular networks, using site-specific ray-tracing propagation data from Frankfurt and Seoul and realistic antenna array patterns. We proposed a greedy heuristic algorithm to solve the network-wide directional link allocation problem, thereby estimating the achievable capacity and coverage of multi-user mm-wave networks. Our results show that inter-cell interference is negligible even in dense multi-user networks with wide antenna beams. Consequently, TDMA throughput is largely insensitive to both antenna model and directionality but is strictly limited by the ratio of air-time sharing among UEs served by a BS. By contrast, SDMA suffers from significant intra-cell interference among the per-UE beams of a serving BS, making it crucial to model real antenna array sidelobes, as ideal sectored beams overestimate the average UE throughput of SDMA by up to $1$~Gbps. Nonetheless, SDMA consistently and significantly outperforms TDMA in average UE throughput, e.g. by up to $2$~Gbps with $8\times8$ arrays. Moreover, our results reveal that larger antenna arrays at the BS benefit SDMA performance by spatially limiting intra-cell interference while compensating for smaller UE arrays. This is an important practical network design insight, as relaxed UE beamforming requirements also simplify mm-wave mobility management. Finally, our SDMA results for HBF with a limited number of BS sub-arrays show that the achieved coverage ratio is a tradeoff between network densification and antenna resource constraints. Our ongoing work is focused on joint scheduling and beam management algorithms that support multi-user provisioning in mobile cellular mm-wave networks.

\ifCLASSOPTIONcaptionsoff
  \newpage
\fi


\bibliographystyle{IEEEtran}
\bibliography{bibliography}

\end{document}